\documentclass[pra,twocolumn,amsmath,amssymb,floatfix,reprint]{revtex4-1}

\usepackage{graphicx}
\usepackage{dcolumn}
\usepackage{bm}
\usepackage[usenames,dvipsnames]{xcolor}
\usepackage{mciteplus}
\usepackage{braket}
\usepackage[normalem]{ulem}
\renewcommand{\emph}{}

\begin{document}

\title{Finite temperature effects in helical
  quantum turbulence }\thanks{Postprint version of the manuscript
  published in Phys. Rev. A {\bf 97}, 043629 (2018).}
\author{Patricio Clark Di Leoni$^{1,2}$\email{clark@df.uba.ar},
  Pablo D.~Mininni$^1$\email{mininni@df.uba.ar}, \&
  Marc E.~Brachet$^3$\email{brachet@physique.ens.fr}}
\affiliation{$^1$Departamento de F\'\i sica, Facultad de Ciencias
    Exactas y Naturales, Universidad de Buenos Aires and IFIBA, CONICET,
    Ciudad Universitaria, 1428 Buenos Aires, Argentina.\\
                  $^2$Department of Physics and INFN, University of
    Rome Tor Vergata, Via della Ricerca Scientifica 1, 00133 Rome, Italy.\\
                  $^3$Laboratoire de Physique Statistique de l'Ecole Normale
    Sup\'erieure associ\'e au CNRS et aux Universit\'es Paris 6 et 7, 24
    Rue Lhomond, 75237 Paris Cedex 05, France.}
\date{\today}

\begin{abstract}
    We perform a study on the evolution of helical quantum turbulence
    at different temperatures by solving numerically the
    Gross-Pitaevskii and the Stochastic Ginzburg-Landau equations,
    using up to $4096^3$ grid points with a pseudospectral method. We
    show that for temperatures close to the critical the fluid
    described by these equations can act as a classical viscous flow,
    with the decay of the incompressible kinetic energy and the
    helicity becoming exponential. The transition from this behavior
    to the one observed at zero temperature is smooth as a function of
    temperature. Moreover, the presence of strong thermal effects can
    inhibit the development of a proper turbulent cascade. We provide
    anzats for the effective viscosity and friction as a function of
    the temperature.
\end{abstract}
\maketitle

\section{Introduction} 
\label{introduction}

In experiments of superfluids and Bose-Einstein condensates (BECs) a
highly disorganized and chaotic behavior, known as quantum turbulence,
can be observed \cite{Vinen02,Henn09,Barenghi14}. At zero temperature quantum
flows are characterized by their lack of viscosity, and by having all
of their vorticity concentrated along vortex filaments with quantized
circulation \cite{Feynman55,Donnelly}. But at finite temperatures
dissipative effects creep in. Landau and Tisza's two fluid model
\cite{Landau}, where a mixture of superfluid and normal fluid coexist
and interact (with the ratio between the two determined by the
temperature), is perhaps the most simple way to represent the finite 
temperature dynamics of superfluids and BECs.

Based on the two fluid model, the Hall-Vinen-Bekarevich-Khalatnikov
(HVBK) model \cite{Hall56,Bekarevich61} adds a term accounting for the
``mutual friction'' between the normal and superfluid components. This
model has been successful in, for example, explaining the Taylor-Couette
instability in liquid helium \cite{Barenghi87}. It has also been used
to study turbulent flows; for example, \citet{Roche09} found that
there is a strong locking between both fluid components and that
both develop a turbulent
cascade, \citet{Shukla15} found the existence of both an inverse and a
forward cascade in the two dimensional case, and shell models based on
the HVBK model were developed and used to study the mutual friction
terms \cite{Wacks11,Boue15}, intermittency \cite{Boue13}, and scaling
exponents \cite{Shukla16}. An alternative to the HVBK model is the
vortex filament model \cite{Schwarz85}, which, as the name implies,
takes the vortex filaments into account explicitly by modeling them as
classical Eulerian vortices of negligible width which evolve under the
Biot-Savart law. As mutual friction can also be added to this model,
it has been used to study quantum turbulence at finite temperatures
\cite{Khomenko15,Khomenko16}. But two important aspects of quantum
turbulence are omitted in these two models. One is the lack of
compressibility effects, and thus, of sound waves. The other is vortex 
reconnection. While in the HVBK model the former is omitted completely,
as the fluid is averaged over volumes larger than the vortex width, in
the vortex filament model it is introduced phenomenologically.

There is another family of models to study finite temperature effects
based on extensions of the Gross-Pitaevskii equation (GPE). At zero or
near zero temperatures the GPE, for which quantized vortices are exact
solutions which can reconnect with no extra {\it ad-hoc} assumptions, is
a very succesful model for BECs \cite{Proukakis08}. Moreover, a
hydrodynamic analogy can be easily obtained from the GPE by means of the
Madelung transformation, and it has been shown that at the larger
scales its turbulent solutions match those of classical turbulence
\cite{Nore97b,Clark17}. There are various ways of generalizing the GPE
for studying finite-temperature effects \cite{Berloff14}. These include
solving the spectrally truncated version of the equations
\cite{Davis01,Connaughton05}, coupling them with a Boltzmann equation
describing the evolution of the thermalized modes as in the
Zaremba-Nikuni-Griffin model \cite{Zaremba99}, or simply adding a
phenomenological dissipation term \cite{Pitaevskii59,Choi98}.  Previous studies of
these models have concentrated on understanding the thermalization
processes \cite{Davis01,Krstulovic11a,Krstulovic11b,Shukla13}, on
investigating single vortex decay
\cite{Kobayashi06,Jackson09,Rooney10,Allen14,Rooney16}, or on modelling
traps with several vortices \cite{Rooney13,Stagg15} in configurations
similar to experiments of BECs \cite{Neely13,Moon15,Kim16,Seo17}.
However, few studies have focused on the properties of the turbulent
motions and on how finite temperature effects come into play in this
regime.

In this context, it is worth noting that the study of quantum turbulence
has garnered much interest in the past years. Two of the main areas of
work have been establishing the differences between classical and
quantum turbulence \cite{Barenghi14,Paoletti08}, and understanding the
dynamics of Kelvin waves \cite{Fonda14,Clark15a}.  The usual picture of
quantum turbulence (see, for example, \cite{Vinen02}) goes by the
following: while at the larger scales the nonlinear energy transfer in
quantum flows is mediated by the interaction between vortices and
reconnection processes \cite{Meichle12}, and the turbulent flow
resembles that of a classical fluid, at scales smaller than the mean
intervortex length Kelvin waves are believed to be the ones responsible
for the energy transfer, thus generating Kelvin wave turbulence
\cite{Kozik04,Lvov10,Boue11,Boue15,Clark15a}. Nonlinear interaction of
Kelvin waves leads to the creation of phonons \cite{Vinen03}, which
are finally responsible for the depletion of incompressible kinetic
energy in quantum turbulence \cite{Nore97a,Clark17}. Additionally,
recently it was shown that at zero temperature helical quantum
turbulence (i.e., for flows with non-zero large-scale helicity)
develops a dual cascade of energy and of helicity reminiscent of the
dual cascade observed in classical helical flows, and that the
emission of phonons also result in the depletion of helicity
\cite{Clark17}. The presence of such a dual cascade, where both 
energy and helicity are being transferred from the larger to the smaller
scales to be finally dissipated, has a strong impact in the evolution
and decay of turbulence. In classical flows, this dual cascade has
received significant attention (see, e.g.,
\cite{Brissaud73,Chen03,Teitelbaum09,Moffatt14}), as well as the effects
of helicity in the evolution and statistical properties of turbulence.
As a result, understanding how helical flows and their dual cascade
are affected by the interaction with the effective thermal dissipation
in finite temperature models will be the first main objective of the
present work.

Indeed, the overall purpose of this paper is to study finite
temperature effects on a helical quantum flow in high resolution
numerical simulations of the truncated Gross-Pitaesvkii equation, with
the thermal states being generated by the Stochastic Ginzburg Landau
method \cite{Berloff14}. Our results show that for high temperatures
the quantum fluid described by this model can behave as a classical
viscous flow, with the decay of energy and of helicity becoming
exponential in time, and with the development of the dual turbulent
cascade being hindered. The transition from the zero to the high
temperature behavior is smooth as a function of the temperature, as
long as the temperature is smaller than the critical. As a second
objective, we will profit from the high spatial resolution of our 
simulations to provide anzats for the effective viscosity as a
function of the temperature. The structure of the paper is as
follows. In Sec.~\ref{model} we outline the physical model used, and
describe the simulations we performed. The main results are presented
in Sec.~\ref{results}. Finally, closing comments are presented in
Sec.~\ref{conclusions}.

\section{The finite temperature model} 
\label{model}

In this section we first present a brief summary of some key concepts
and definitions of the zero temperature model (the GPE), used in this
work as the starting point for the finite temperature model. Then, we
explain how to generate finite temperature states using the Stochastic
Ginzburg Landau equation (SGLE), following the method outlined in
\cite{Krstulovic11a,Krstulovic11b}, and how to use these states in
quantum turbulence simulations solving the GPE.  Finally, we give
details of a large number of high resolution simulations performed for
the present study.

\subsection{The Gross-Pitaevskii equation} 

At zero (or near zero) temperatures, a field of weakly interacting
bosons can be appropriately described by the GPE,
\begin{equation}
    i \hbar \frac{\partial \Psi}{\partial t}
    =
    - \frac{\hbar^2}{2m} \nabla^2 \Psi 
    + g \vert \Psi \vert^2 \Psi,
    \label{gpe}
\end{equation}
where $\Psi$ is the wavefunction of the condensate, $m$ is the mass of
the bosons, and $g$ is proportional to the bosons scattering
length. The GPE conserves the total energy
\begin{equation}
    E = \int_V dV \left( \frac{\hbar^2}{2m} \vert \nabla \Psi \vert^2 + \frac{g}{2}
    \vert \Psi \vert^4 \right),
\end{equation}
the momentum
\begin{equation}
    {\bf P} = \int_V dV \frac{i\hbar}{2} \left( \Psi \nabla \bar{\Psi} -
    \bar{\Psi} \nabla \Psi \right),
\end{equation}
(where the overbar denotes complex conjugate), and the total number of
particles
\begin{equation}
    \mathcal{N} = \int_V dV \vert\Psi\vert^2 .
\end{equation}

A hydrodynamical description of the flow can be recovered via the
Madelung transformation 
\begin{equation}
    \Psi ({\bf r},t) = \sqrt{\frac{\rho ({\bf r},t)}{m}} e^{i m \phi
    ({\bf r},t)/\hbar},
\end{equation}
where $\rho({\bf r},t)$ is the fluid mass density, and $\phi({\bf r},t)$ is
the velocity potential. Applying this transformation to the GPE yields
the equations for an ideal barotropic fluid plus an extra term with
the gradient of the so-called quantum pressure. This hydrodynamical
description is useful to separate the total energy into different
components \cite{Nore97a}. These are respectively the kinetic energy
\begin{equation}
    E_k = \int_V dV \frac12 \rho \vert {\bf v} \vert^2,
\end{equation}
(which in turn can be separated into an incompressible component
$E^i_k$ and a compressible one $E^c_k$ using a Helmholtz decomposition
of the velocity field), the quantum energy
\begin{equation}
    E_q = \int_V dV \frac{\hbar^2}{2 m^2} (\nabla \sqrt{\rho})^2,
\end{equation}
and the internal (or potential) energy
\begin{equation}
    E_p = \int_V dV \frac{g}{2 m^2} \rho^2.
\end{equation}

\begin{figure}
    \centering
    \includegraphics[width=8.5cm]{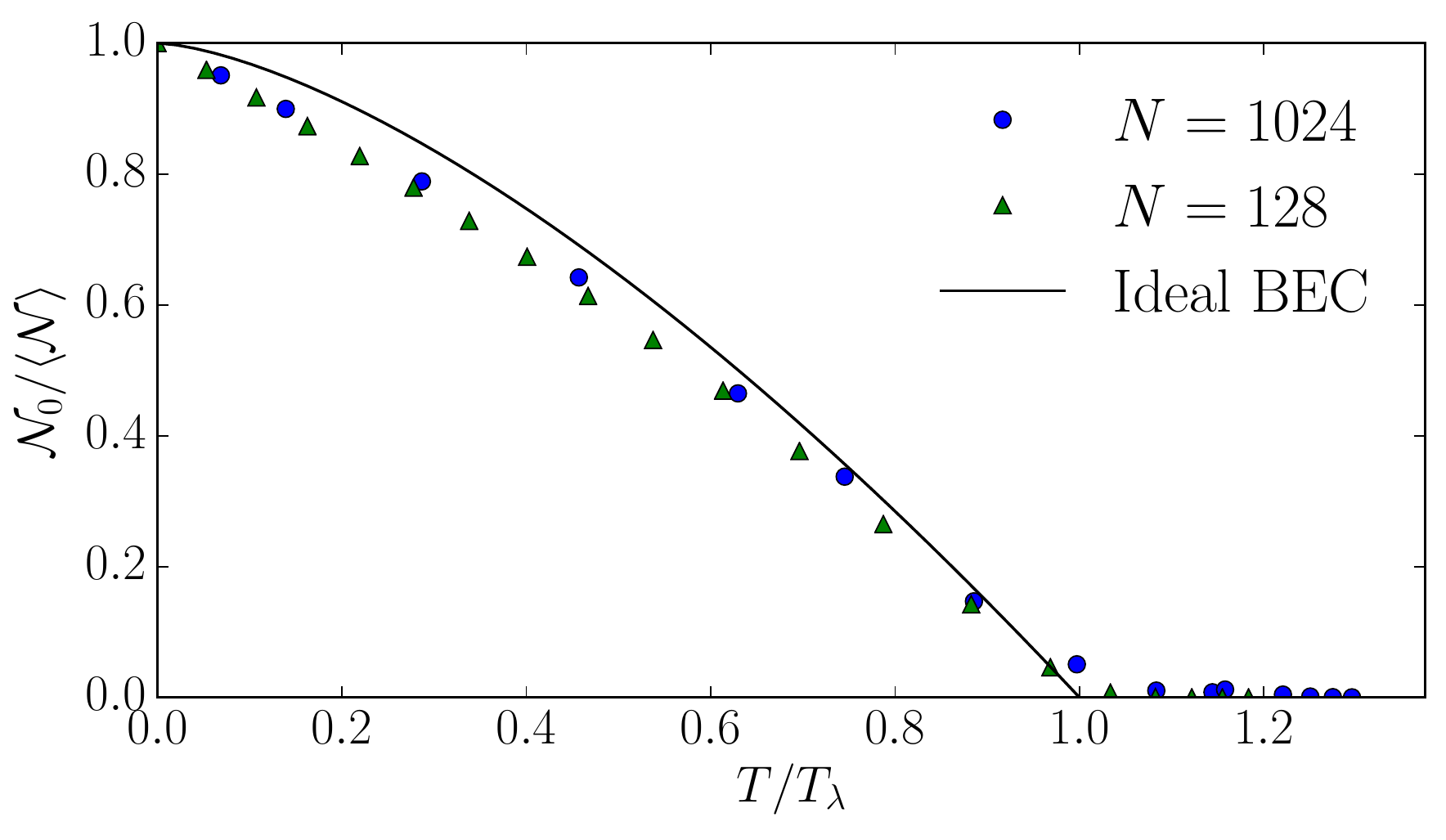}
    \caption{{\it (Color online)} Condensate fraction
    $\mathcal{N}_0/\langle\mathcal{N}\rangle$ of constant total density
    scans as a function of the temperature at two different spatial
    resolutions. The simulations with $N^3$ grid points, with $N=1024$,
    are marked with (blue) circles, while the simulations with $N=128$
    are marked with (green) triangles. The solid black line indicates
    the usual ideal BEC theory prediction for the condensate fraction as a
    function of temperature.}
    \label{condf}
\end{figure}

By linearising Eq.~\eqref{gpe} around $\Psi = \Psi_0$ (constant), one
can obtain the Bogoliubov dispersion relation 
$\omega_B(k) = c k (1 + \xi^2 k^2/2)^{1/2}$, 
where $c=[g \vert \Psi_0 \vert^2/m]^{1/2}$ is the speed of sound and
$\xi = [\hbar^2 / (2m \vert \Psi_0 \vert^2 g) ]^{1/2}$ is the healing
length. The GPE can also sustain Kelvin waves, which are helical
perturbations that travel along the quantum vortices. As stated in 
Sec.~\ref{introduction}, Kelvin waves play a major role in zero
temperature quantum turbulence, where they are responsible for the
energy transfer at scales smaller than the intervortex distance.

\begin{figure}
    \centering
    \includegraphics[width=8.5cm]{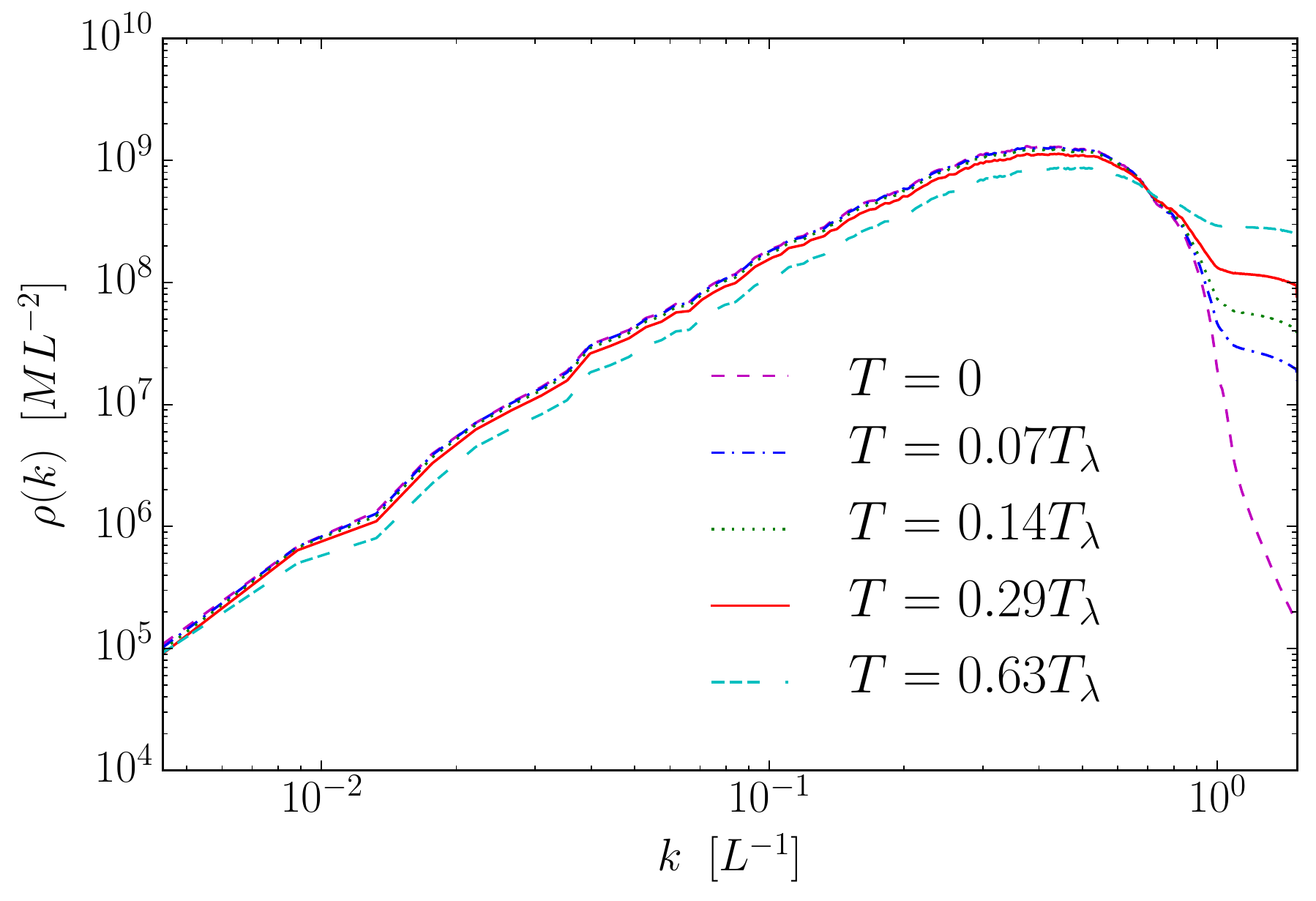}
    \includegraphics[width=8.5cm]{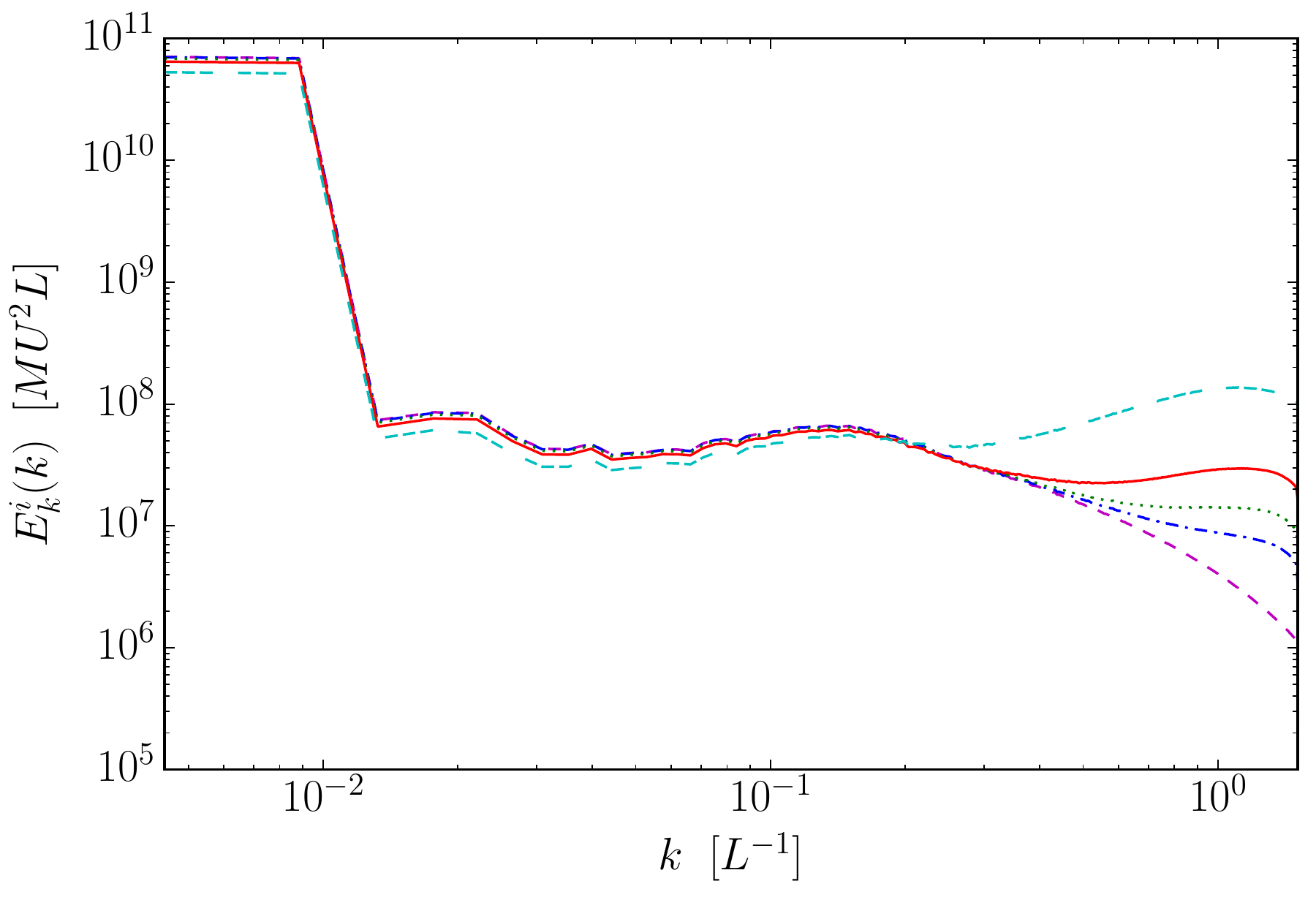}
    \includegraphics[width=8.5cm]{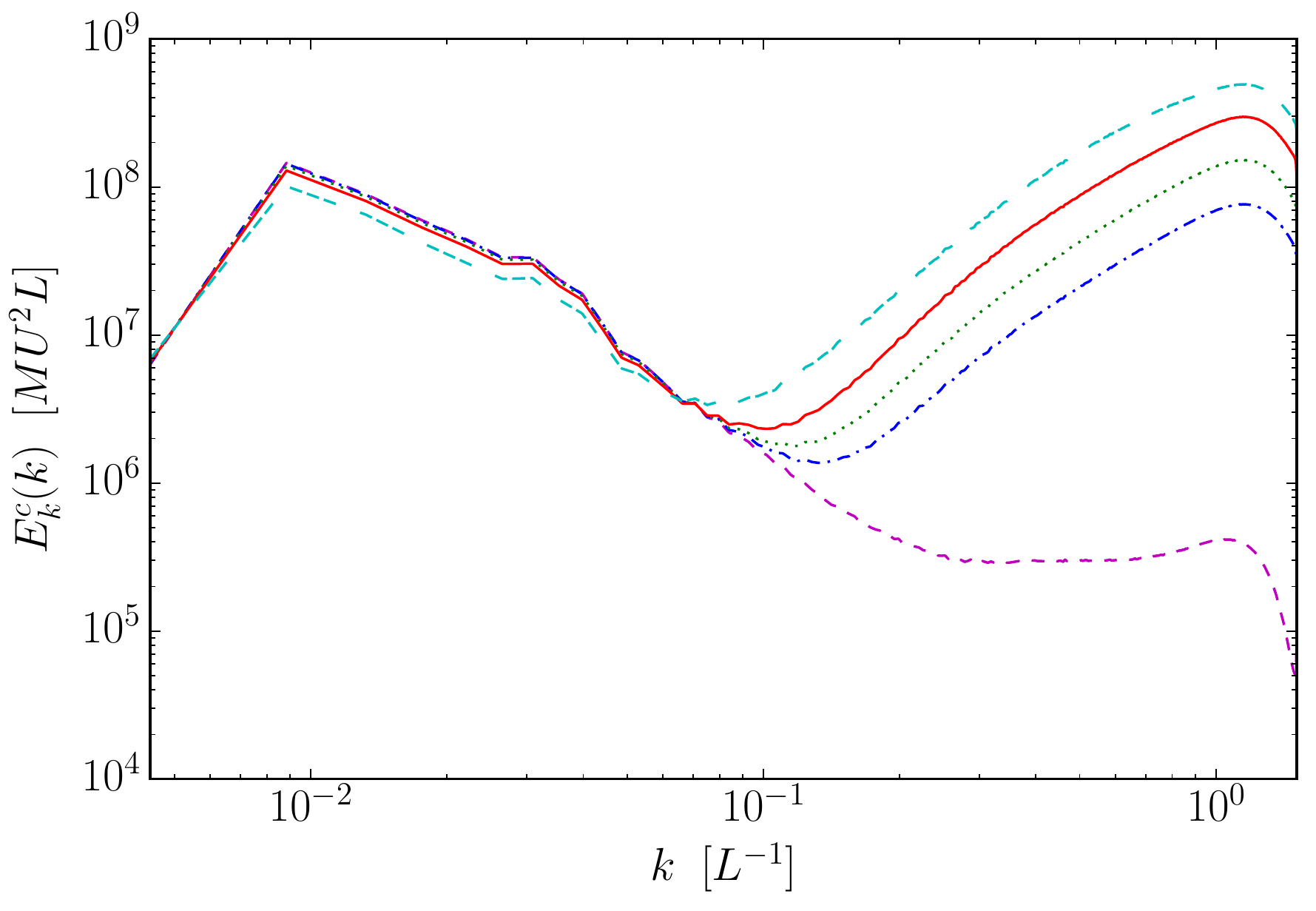}
    \caption{{\it (Color online)} From top to bottom, spectra of mass
      fluctuations, of the incompressible kinetic energy, and of the
      compressible kinetic energy, for several initial conditions of
      the GPE at different temperatures. All simulations have $N=1024$
      linear spatial resolution.}
    \label{massk}
\end{figure}

One last aspect of the GPE dynamics of relevance for this work is the
concept of helicity. In classical fluids the helicity is defined as
\begin{equation}
    H = \int_V dV {\bm v} \cdot {\bm \omega} ,
\end{equation}
where ${\bf \omega}$ stands for the vorticity field. Helicity is a
measure of the mean alignment between velocity and vorticity (and
thus, of the depletion of nonlinearities), of the topological complexity
of vorticity field lines, as well as a measure  of the departure of
mirror-symmetry of the flow
\cite{Moffatt69,Moffatt92,Moffatt92b,Moffatt14}. In classical
turbulence the presence of helicity in a turbulent flow can have
multiple consequences, such as the depletion of the nonlinearities and
energy transfers \cite{Kraichnan73}, the slowing down of the onset of
dissipation \cite{Andre77}, and it can even affect the evolution of
convective storms \cite{Lilly86}. It has also been show that the
helicity, just like the energy, develops a turbulent cascade where it
is transferred from the larger to the smaller scales
\cite{Brissaud73}. Moreover, the form of the cascade implies that it
is a {\it dual} cascade, meaning that both energy and helicity have
simultaneously non-zero transfer rates in the inertial range. In
quantum fluids, Kelvin waves are helical and thus $H$ could in
principle be used as a proxy to quantify the excitation of Kelvin 
waves at small scales. However, both ${\bf v}$ and ${\bf \omega}$ are
singular along the vortex lines of a quantum fluid, where all the
vorticity is concentrated. To overcome this problem, many authors have
chosen to work with a definition of helicity based on its topological
interpretation \cite{Scheeler14}. These geometric decompositions can
result in zero net helicity \cite{Hanninen16} but recover a classical
non-zero value at large-scales \cite{Salman17,Kedia17}. Other authors
have chosen to work with filtered fields \cite{Zuccher15}. Here we
will use the {\it regularized} helicity introduced in \cite{Clark16},
where the velocity field is regularized  before being used to compute
$H$. This method was shown to give results compatible with other
methods in the literature to estimate the helicity of a quantum flow,
and was used successfully to study helical quantum turbulence at zero
temperature in massive numerical simulations in \cite{Clark17}, where
the existence of a dual cascade of energy and helicity was confirmed
for the quantum case.

\begin{figure}
    \centering
    \includegraphics[height=7cm]{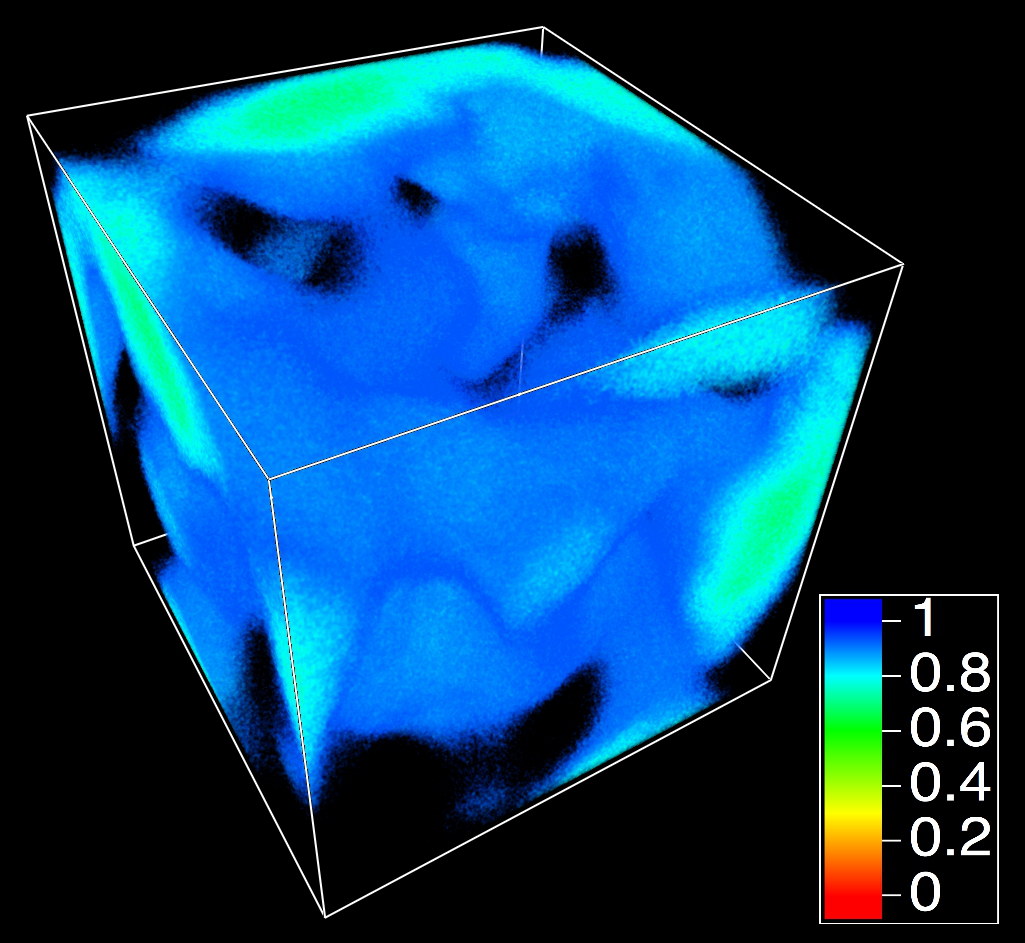}
    \caption{{\it (Color online)} Volume rendering of the density field
    for the simulation with $N=4096$ and $T=0.64T_\lambda$. Similar to
    the zero temperature ABC flow \cite{Clark17}, large structures and
    regions of quiescencenot present in the initial conditions 
    are spontaneously formed within the flow. At the large scales, the
    flow resembles the structure of a classical ABC flow. The
    possibility of seeing these large scale structures formed by the
    quantized vortices (the smallest structures in the flow) in such
    detail is in part due to the large scale separation, product of the
    high resolution used in the simulation. At low resolution there is
    not enough scale separation between the large scales and the
    thermal fluctuations for such a structures to develop.}
    \label{fullbox_dvr}
\end{figure}

\subsection{The Stochastic Ginzburg Landau equation} 

The spatially truncated version of different conservative systems of
partial differential equations can achive, after long time integration,
states of thermodynamic equilibrium known as thermalized states where
energy is equipartitioned among all the possible spatial modes
\cite{Lee52,Kraichnan89}. A common way to truncate a system is via a
Galerkin projector. Given a Fourier series expansion of the
wavefunction
\begin{equation}
    \Psi({\bm r},t) = \sum^{\infty}_{k=-\infty} \hat{\Psi}_{\bm
    k}(t) e^{i{{\bm k}\cdot \bm{r}}},
\end{equation}
where $\hat\Psi_{\bm k}$ are the Fourier coefficients and ${\bm k}$
are the wavevectors, the projector has the form
\begin{equation}
    {P}_{k_G} [ \Psi ({\bm r},t)] = \sum_{|k|\leq k_G} \hat{\Psi}_{\bm
    k}(t) e^{i{{\bm k}\cdot \bm{r}}} .
    \label{galerkin}
\end{equation}
Applying it to Eq.~\eqref{gpe} would give the so-called Fourier (or
Galerkin) truncated version of the GPE.

\begin{figure}
    \centering
    \includegraphics[height=7cm]{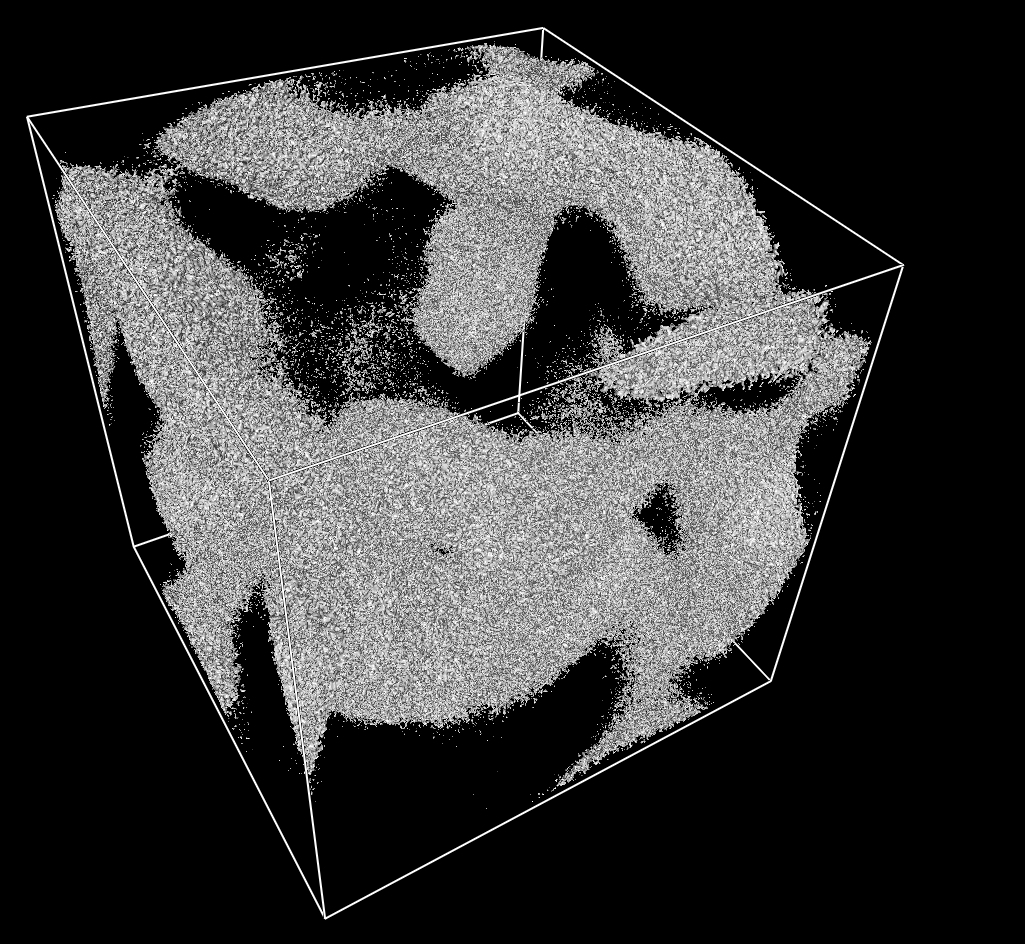}
    \caption{{\it (Color online)} Isosurfaces of the density field for
      the simulation with $N=4096$ and $T=0.64T_\lambda$. Contrary to
      the zero temperature case \cite{Clark17}, it is not possible to
      discern individual vortices now. But their presence in the flow
      is still evident when looking at the fine-grain structures.
      Also, the formation of the vortex bundles observed in the
        zero temperature case is hampered in this case.}
    \label{fullbox_iso}
\end{figure}

The studies of \citet{Davis01} and of \citet{Connaughton05} showed that
if the Fourier truncated version of the GPE is integrated for long
enough, the system indeed reaches a thermodynamic equilibrium. The
statistical properties of this state are given by the microcanonical
ensamble defined with fixed energy $E$, momentum ${\bf P}$, and number
of particles $\mathcal{N}$. Moreover, if $E$ is varied, a phase
transition akin to that of BECs can be observed, where the
zero-wavenumber $A_0 = \langle\Psi\rangle$ mode becomes equal to zero
for finite $E$. But there are two problems with generating thermal
states in this way.  One is that the truncated GPE takes a very long
time to converge to the equilibrium state, making it computationally
expensive. The other is that the temperature is not easily accessed nor
controlled in this way, given the complicated expression for the entropy
in the microcanonical state of the the system. In order to overcome
these problems, \citet{Krstulovic11a,Krstulovic11b} suggested using a
Langevin process to generate grand-canonical states with distribution
probability $\mathbb{P}_{\rm st}$ given by a Boltzmann weight
$\mathbb{P}_{\rm st}=e^{-\beta F}/\mathcal{Z}$, where $\mathcal{Z}$
denotes the grand partition function and \begin{equation} F = E - \mu
\mathcal{N} - {\bf W}\cdot{\bf P}, \end{equation} is a free energy with
$\beta$ the inverse temperature, $\mu$ is the chemical potential, and
${\bf W}$ is related to the counterflow velocity. These grand-canonical
states are faster to generate than microcanonical states,  and give easy
access and control of the temperature in the equilibrium.

\begin{figure}
    \centering
    \includegraphics[width=8.5cm]{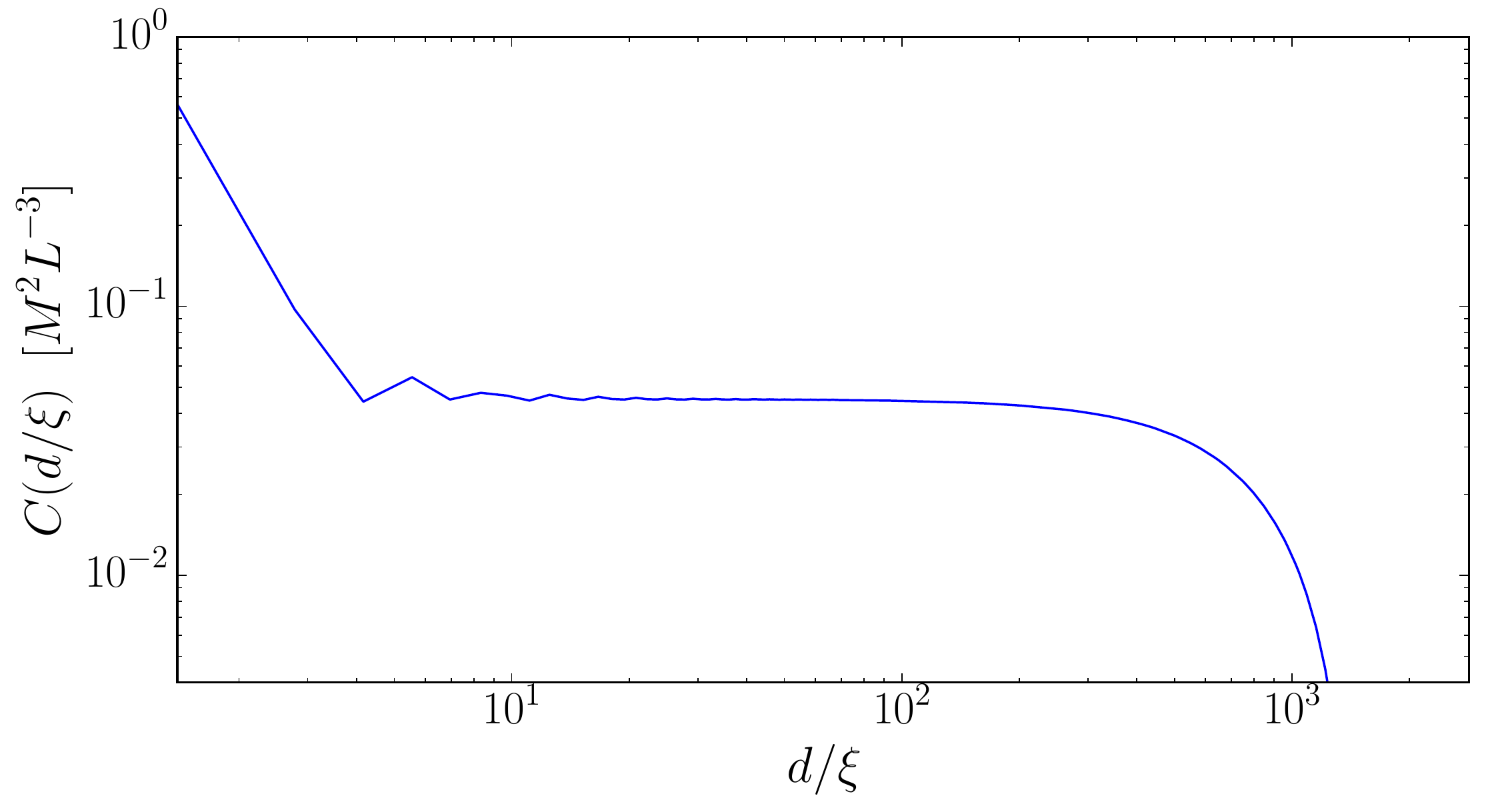}
    \caption{{\it (Color online)} Mass density correlation function
      $C(d) = \left< (\rho({\bf x} +d\hat{x}) - \rho_0)(\rho({\bf x})
        - \rho_0) \right>$ for the simulation with $N=4096$ and
      $T=0.64T_\lambda$ at $t\approx 1$, with the displacement $d$
      normalized in units of the healing length $\xi$. Note the
      slow decay of the correlation up to distances 
      $\approx 10^3 \xi$.}
    \label{fullbox_corr}
\end{figure}

The Langevin process that generates these states has a Ginzburg-Landau
equation of the type
\begin{gather}
    \hbar \frac{\partial A_{\bf k}}{\partial t} = - \frac{\partial
    F}{\partial A^*_{\bf k}}
    + \sqrt{\frac{2 \hbar}{\beta}} \hat{\xi} ({\bf k},t), \label{sglespec1}
    \\
    \langle \xi ({\bf r},t) \bar{\xi} ({\bf r}',t')\rangle = \delta(t-t')
    \delta({\bf r} - {\bf r}'),\label{sglespec2}
\end{gather}
where $A_{\bf k}$ are the Fourier modes of the wavefunction, and
$\hat{\xi}({\bf k},t)$ is the Fourier transform of the Gaussian
delta-correlated noise $\xi({\bf r},t)$. In \cite{Krstulovic11b} it is
shown that the stationary probability of the solutions of
Eq.~\eqref{sglespec1} is indeed $\mathbb{P}_{\rm st}$. Thus, the
grand-canonical states are simply generated by integrating the
Langevin Eq.~\eqref{sglespec1} in time until statistical convergence
is obtained.

In physical space, the Langevin equation reads
\begin{align}
    \hbar \frac{\partial \Psi}{\partial t} = &\left[ \frac{\hbar^2}{2m}
    \nabla^2 \Psi + \mu \Psi - g \vert \Psi \vert^2 \Psi - i \hbar {\bf
    W}\cdot \nabla \Psi \right]
    \nonumber
    \\
    &+ \sqrt{\frac{2\hbar}{\beta}} \xi .
    \label{sgle}
\end{align}
This equation will be referred to as the Stochastic Ginzburg-Landau
equation (SGLE). The chemical potential $\mu$ controls the total number
of particles $\mathcal{N}$. Different solutions obtained by varying
$\beta$ will have different ratios of condensed fraction
$|A_0|^2/\mathcal{N} = \mathcal{N}_0/\langle\mathcal{N}\rangle$, except below a
critical $\beta$ (or, in terms of temperature, above the transition
temperature $T_\lambda$) where this ratio will be equal to zero.

The thermal states obtained from the SGLE can then be fed to the GPE,
in combination with an initial condition for the large-scale flow, to
simulate a quantum turbulent flow at finite temperature. The total
initial condition for the GPE $\Psi$ then has the form
\begin{equation}
    \Psi = \Psi_{\mathrm{flow}} \times \Psi_{\mathrm{SGLE}},
    \label{inipsi}
\end{equation}
where $\Psi_{\mathrm{flow}}$ is an initial wavefunction describing the
flow, and $\Psi_{\mathrm{SGLE}}$ is a thermal solution of the SGLE
which gives account of the occupation numbers of the different
energy levels in the thermal state at a given temperature.

Although for simplicity the projector defined in Eq.~\eqref{galerkin} is
not explicity written in Eqs.~\eqref{gpe} and \eqref{sgle}, in the
following we will indeed solve the truncated versions of each equation.
It is also worth noting that {\it every} time one solves a system of
partial differential equations numerically, one is actually solving for
truncated equations. Depending on the numerical method used for spatial
discretization, the integration can preserve the conservation properties
of the truncated system or not. The method used here, and described
next, preserves all quantities conserved by the Galerkin truncated
Eqs.~\eqref{gpe} and \eqref{sgle} in the continuum-time case (i.e.,
before time discretization).

\begin{figure}
    \centering
    \includegraphics[width=8.5cm]{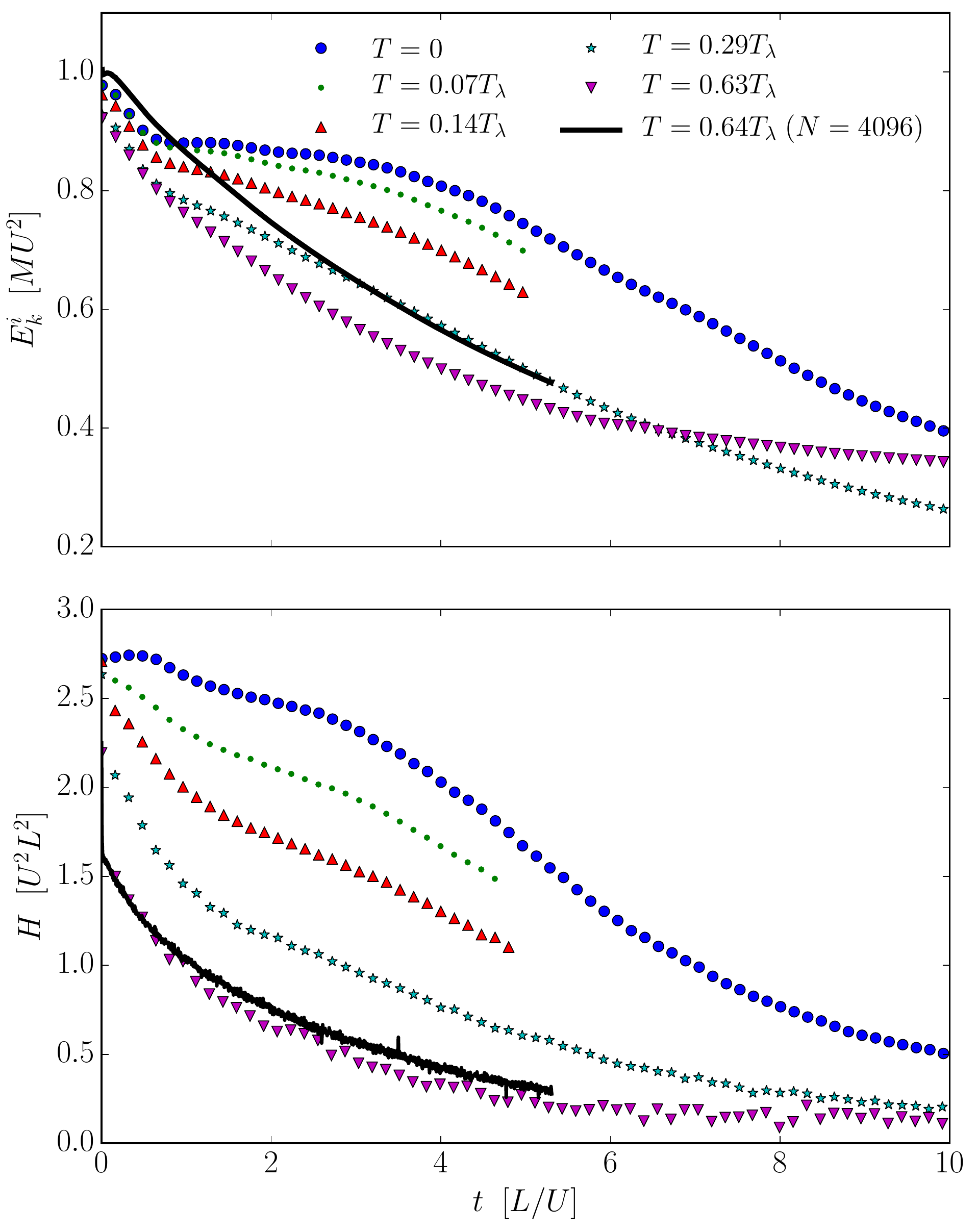}
    \caption{{\it (Color online)} Evolution of the incompressible
      kinetic energy (top) and of the helicity (bottom) as function of
      time at different temperatures. All simulations have $N=1024$ linear
      resolution, except the the one indicated with the solid black
      line which has $N=4096$. The early ``inviscid-like'' behavior
      seen at low temperature, in which energy and helicity remain
      approximately constant, is lost as the temperature is increased.}
    \label{eht}
\end{figure}

\begin{figure}
    \centering
    \includegraphics[width=8.5cm]{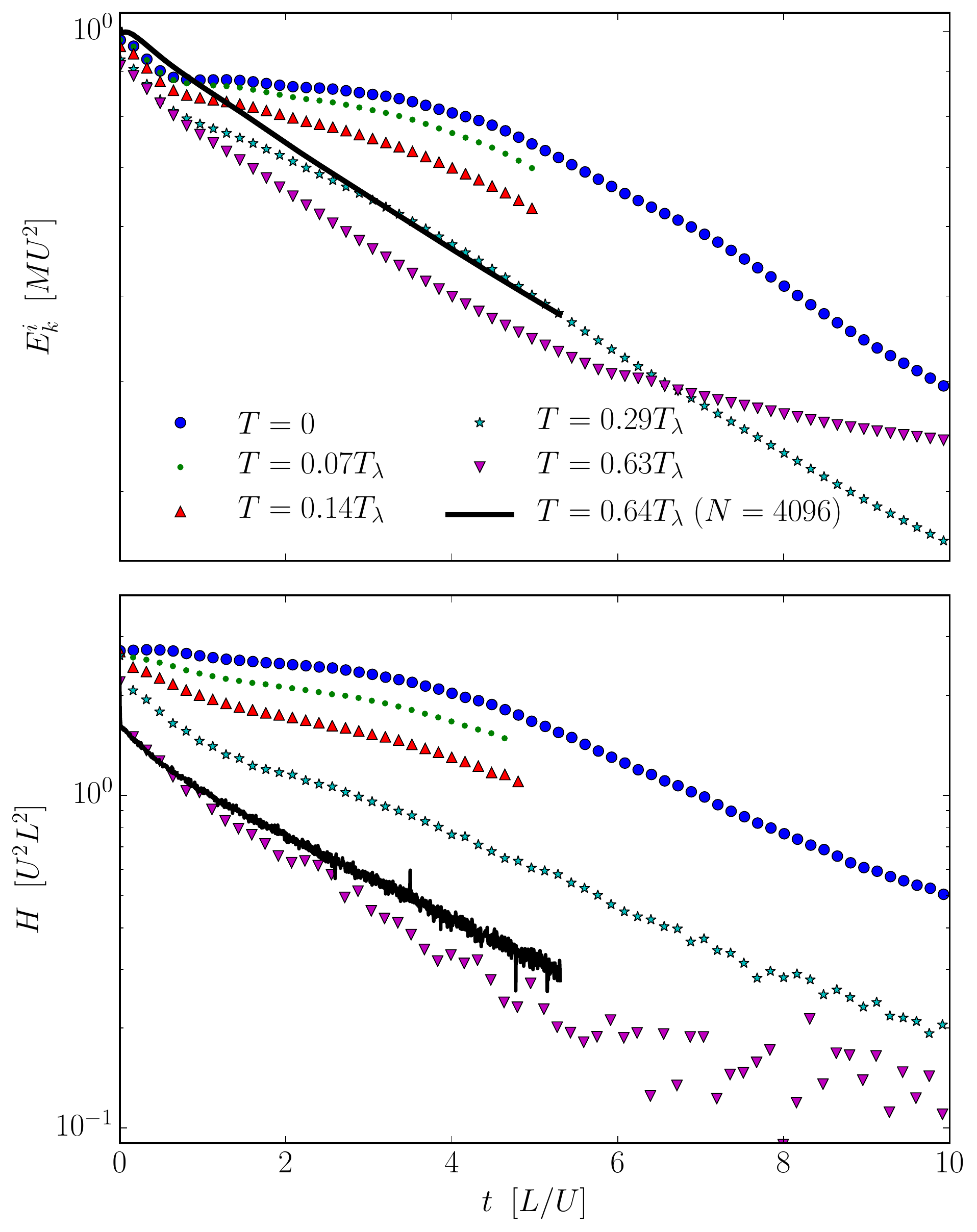}
    \caption{{\it (Color online)} Evolution of the incompressible
      kinetic energy (top) and of the helicity (bottom) as a function
      of time and at different temperatures in semi-logarithmic
      scale. All simulations have $N=1024$, except the one indicated
      with the solid black line which has $N=4096$. At the highest
      temperatures quantities decay exponentially in time.}
    \label{ehtlog}
\end{figure}

\subsection{Initial conditions and numerical simulations} 

To solve numerically Eqs.~\eqref{gpe} and \eqref{sgle} in three
dimensions we used GHOST \cite{Mininni11}, which uses a pseudospectral
method combined with a fourth order Runge-Kutta scheme to solve
Eq.~\eqref{gpe}, and an implicit Euler scheme to solve
Eq.~\eqref{sgle}. Boundary conditions are periodic, each side of the
simulation box is of size $2\pi L$ (where $L$ is a characteristic
scale of the flow), and the ``2/3 rule'' is used for dealising. A
hybrid OpenMP-MPI scheme is used for the parallelization. Multiple
simulations were done at three different spatial resolutions $N^3$,
with linear resolutions $N=128$, $N=1024$ and $N=4096$. In all cases,
the speed of sound is chosen to be twice the characteristic flow
velocity. For the simulation at the largest resolution ($N=4096$), a
total of 8192 processors were used with 4096 MPI jobs and 2 threads
per MPI job, and over 16 million CPU hours were used for the
integration of this simulation.

All quantities are made dimensionless using a characteristic length, a
speed, and a mass. Quantities with units can be determined at any time
by doing $L=L'/(2\pi)$, $U=c'/2$, and $M=M'/(2\pi)^3$, where
$L'$ is the characteristic length of the physical system, $c'$ is the
speed of sound, and $M'$ is the fluid or gas mass (note all primed
quantities have units). With this choice, the length of the simulation
domain is equal to $2\pi L$, the speed of sound $c$ is equal to $2U$,
and the mean density $\rho_0$ is equal to $1 \, M/L^3$. The healing
length $\xi$ is such that $k_{\max}\xi=1.5$, where $k_{\max} = N/3$ (in units  
of $1/L$) is the largest resolved wavenumber in each simulation (the
equivalent of $k_G$ in Eq.~\eqref{galerkin}). In the highest resolved
simulation, the healing length then is $\xi \approx 0.0011 L$.  As a
referece, in superfluid $^4$He experiments the characteric system
size is $L' \approx 10^{-2}$ m, the speed of sound is $c'\approx 230$
m/s, the fluid density is $\approx 125$ kg/m$^3$ (thus 
$M' \approx 1.25 \times 10^{-4}$ kg), and the healing length is 
$\xi' \approx 10^{-8} \, \textrm{m} \approx 10^{-6} L$
\cite{Barenghi14}. The insufficient scale separation of our highest
resolved simulation (even with the massive resolution considered) is
however much better suited for comparisons with BECs. In this case
$L' \approx 10^{-4}$ m, $c'\approx 2\times10^{-3}$ m/s, and 
$\xi \approx 5 \times 10^{-7} \, \textrm{m} \approx 0.005 L$  
\cite{White14}. For the sake of simplicity, in the following all
quantities are quoted using $L=M=U=1$, units can be added later
using the procedure explained above. Finally, temperatures in the
following will be always expressed explicitly in units of the
transition temperature $T_\lambda$. More details on how units can be
handled in GPE and SGLE simulations can be found in 
\cite{Nore97a,Krstulovic11a,Krstulovic11b}.

Simulations with $N=128$ and with $N=1024$ were performed at different
temperatures, while only one simulation at a fixed temperature was
performed at $N=4096$. All simulations were performed with no
counterflow, so ${\bf W}$ in Eq.~\eqref{sgle} is always set to zero, and
the normal and superfluid components are in all cases in perfect coflow.

\begin{figure}
    \centering
    \includegraphics[width=8.5cm]{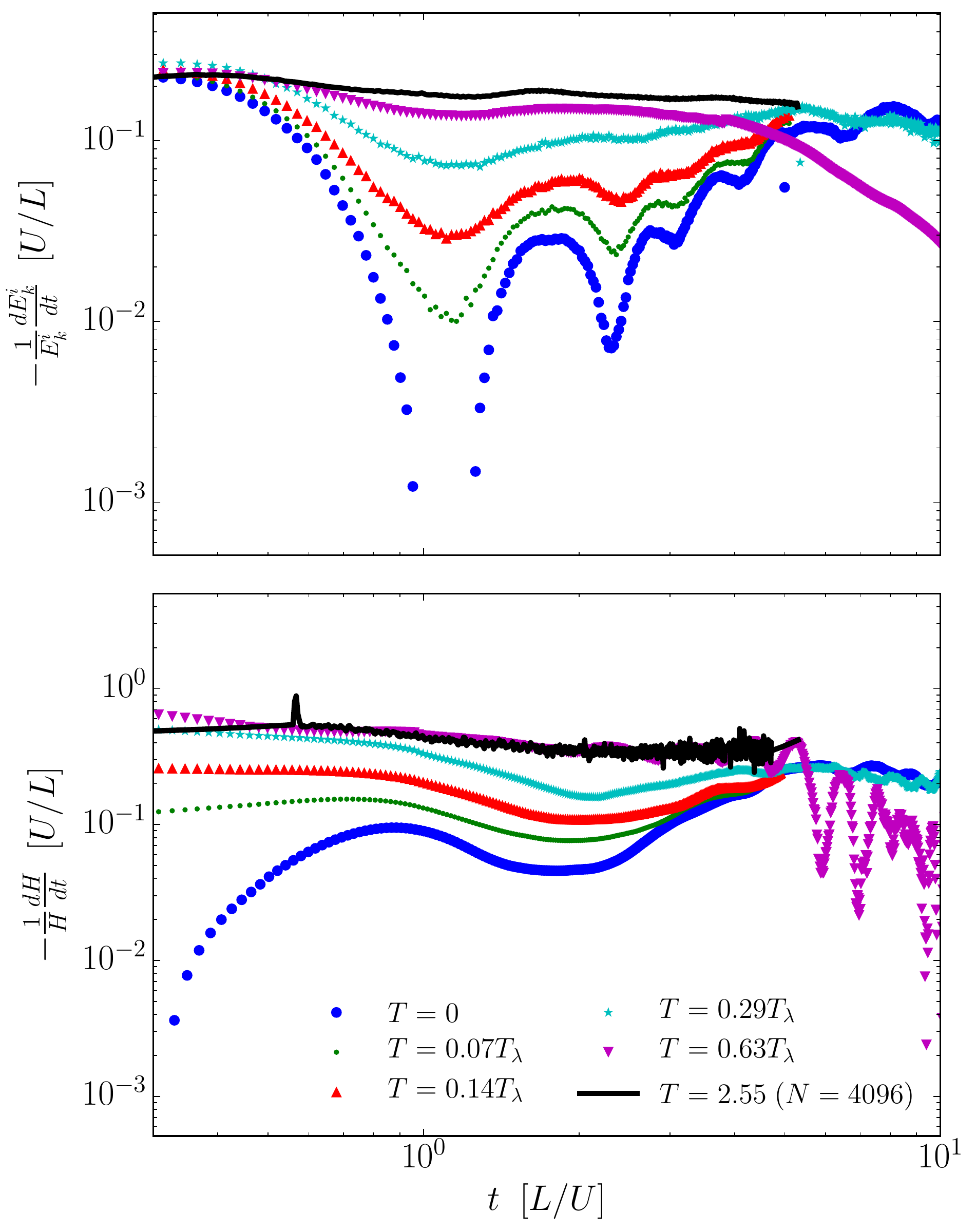}
    \caption{{\it (Color online)} Confirmation of exponential decay
      for the high temperature simulations, akin to that of a viscous
      classical fluid: incompressible kinetic energy exponential decay
      rate $-(1/E^i_k) dE^i_k/dt$ for different temperatures in
      $N=1024$ runs (top), and same for the helicity $-(1/H) dH/dt$
      (bottom). While simulations at low temperature display
      oscillations at early times and growth at late times, at the
      highest temperatures these quantities remain approximately
      constant for long times, allowing us to estimate an exponential
      decay rate.}
    \label{exponential}
\end{figure}

In order to get a helical flow at large-scales, for the flow initial
conditions $\Psi_{\mathrm{flow}}$ we used a superposition of two
quantum Arnold-Beltrami-Childress (ABC) flows \cite{Clark17}. The
velocity field is a superposition of an ABC flow at $k=1$ and of an
ABC flow at $k=2$: 
${\bf v}_{\rm ABC}={\bf v}_{\rm ABC}^{(1)}+{\bf v}_{\rm ABC}^{(2)}$,
with
\begin{eqnarray}
{\bf v}_{\rm ABC}^{(k)} = & \left[ B \cos(k y) + C \sin(k z) \right]
{\bf i}
+ \left[ C \cos(k z) + \right. \nonumber \\
  {}& \left. A \sin(k x)  \right] {\bf j} +
    \left[ A \cos(k x) + B \sin(k y) \right] {\bf k}
\label{ABC}
\end{eqnarray}
with $(A,B,C)=(0.9,1,1.1)/\sqrt{3}$, and where ${\bf i}$, ${\bf
j}$, and ${\bf k}$ are the three Cartesian vectors. The wavefunction
that generates this flow after a Madelung transformation is obtained by
the following procedure, detailed in \cite{Clark16}. First, we set
$\Psi_{\rm flow}=\Psi_{\rm ABC}^{(1)} \times \Psi_{\rm ABC}^{(2)}$,
with
$\Psi_{\rm ABC}^{(k)}= \Psi_{A,k}^{x,y,z}  \times \Psi_{B,k}^{y,z,x}  
    \times \Psi_{C,k}^{z,x,y}$, and with
$\Psi_{A,k}^{x,y,z} =\exp\{i [A \sin(k x)\,m/\hbar] y
    +i [A \cos(k x)\,m/\hbar] z\}$, where $[a]$ stands for the nearest
integer to $a$. In order to minimize the amount of energy in acoustic
modes at the initial condition, we then evolve $\Psi_{\mathrm{flow}}$
using the advected real Ginzburg-Landau equation (ARGLE), whose
stationary solutions are solutions of the GPE with minimal amount of
phonons. The ARGLE explicitly reads
\begin{align}
\partial_t \Psi =&  \frac{\hbar}{2 m} \nabla^2 \Psi 
    +(\frac{g\rho_0}{m}-g|\Psi |^2
    -\frac{m {\bf v}_{\rm ABC}^2}{2 \hbar})\Psi \nonumber \\
& -i {\bf v}_{\rm ABC} \cdot \nabla  \Psi.
 \end{align}
More information on the ARGLE can be found in \cite{Nore97a}, while
the details of the quantum ABC flow are discussed in
\cite{Clark16,Clark17}. The resulting flow has maximal helicity, and
was used in \cite{Clark17} to study helical quantum turbulence at zero
temperature.

Once $\Psi_{\mathrm{flow}}$ has been computed, we solve
Eq.~\eqref{sgle} to obtain a thermal solution at a given temperature,
and finally we compute the initial conditions for the GPE using
Eq.~\eqref{inipsi}.

\begin{figure}
    \centering
    \includegraphics[width=8.5cm]{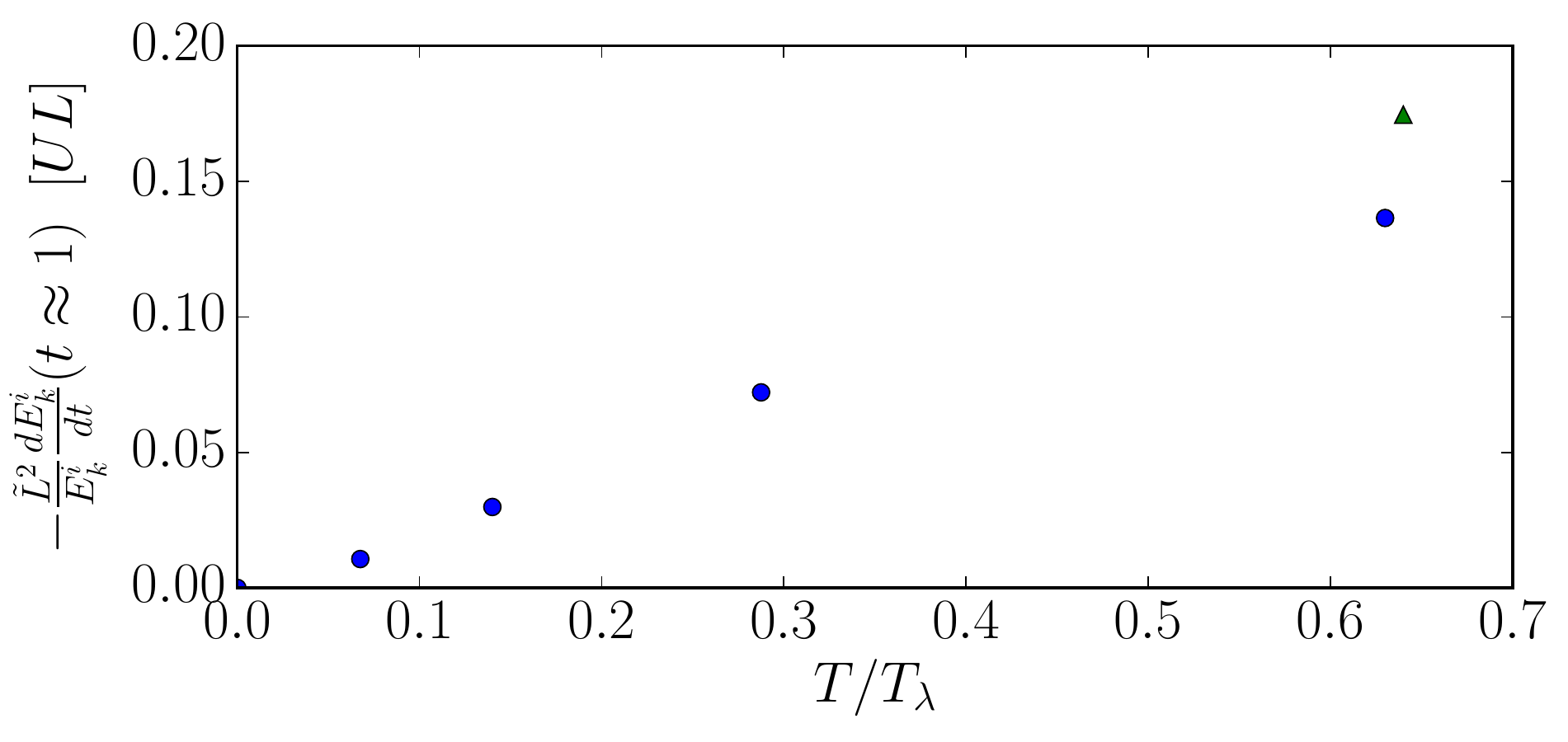}
    \includegraphics[width=8.5cm]{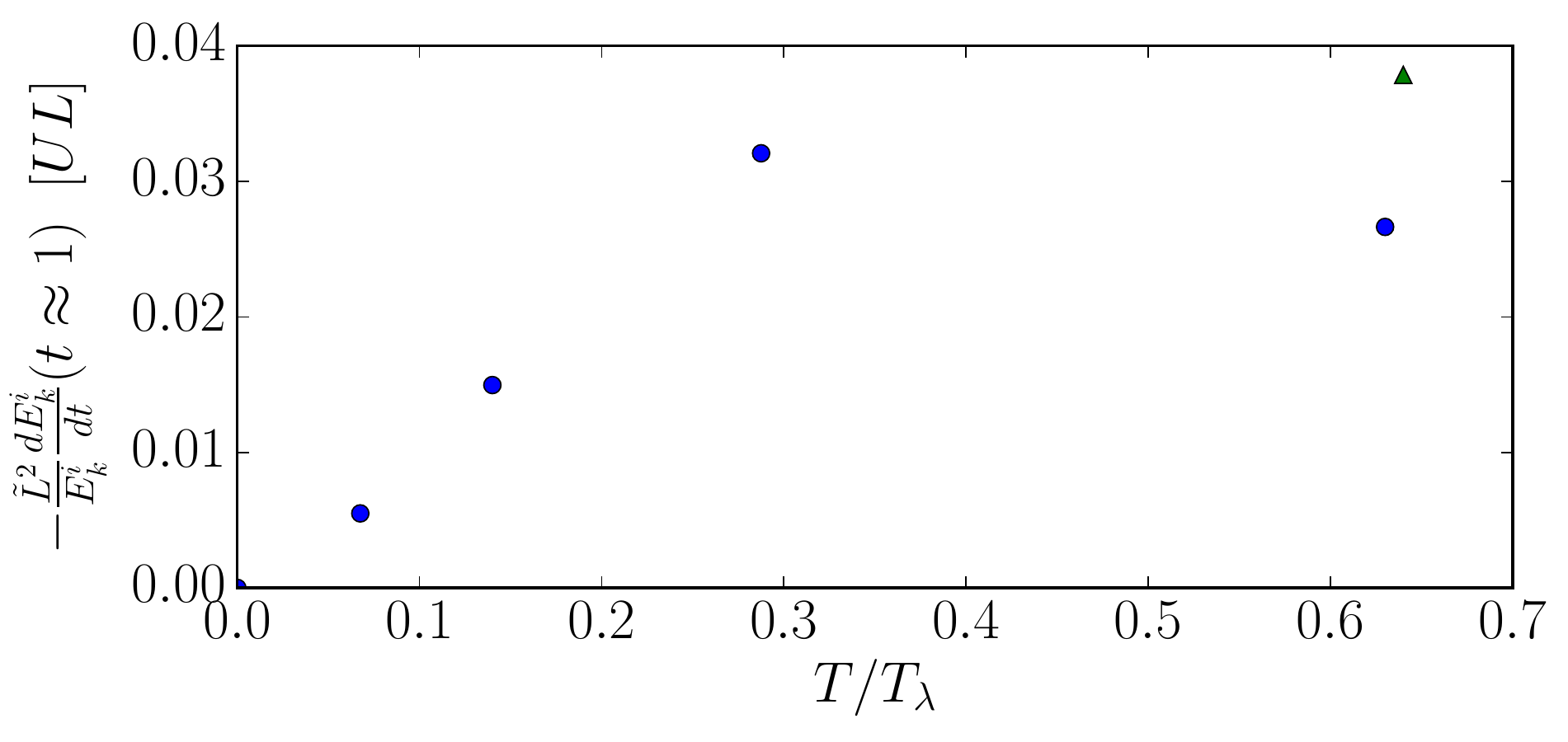}
    \caption{{\it (Color online)} Estimation of the effective
      viscosity from the energy decay rate, 
      $\nu_\textrm{eff} = -(\tilde{L}^2/E^i_k) dE^i_k/dt$ in the vicinity of
      $t\approx 1$, as a function of the temperature. Two choices for
      the characteristic scale are shown: the lengthscale of the
      initial ABC flow $\tilde{L}=L_0$ (top), and the correlation length of
      the incompressible velocity field $\tilde{L}=L_i$ in the vicinity of
      $t\approx 1$ (bottom). The (blue) circles indicate the
      simulations with $N=1024$, and the (green) triangle the
      simulation with $N=4096$.}
    \label{scaling}
\end{figure}

\section{Numerical results} 
\label{results}

\subsection{Temperature scans} 

\begin{figure}
    \centering
    \includegraphics[width=8.5cm]{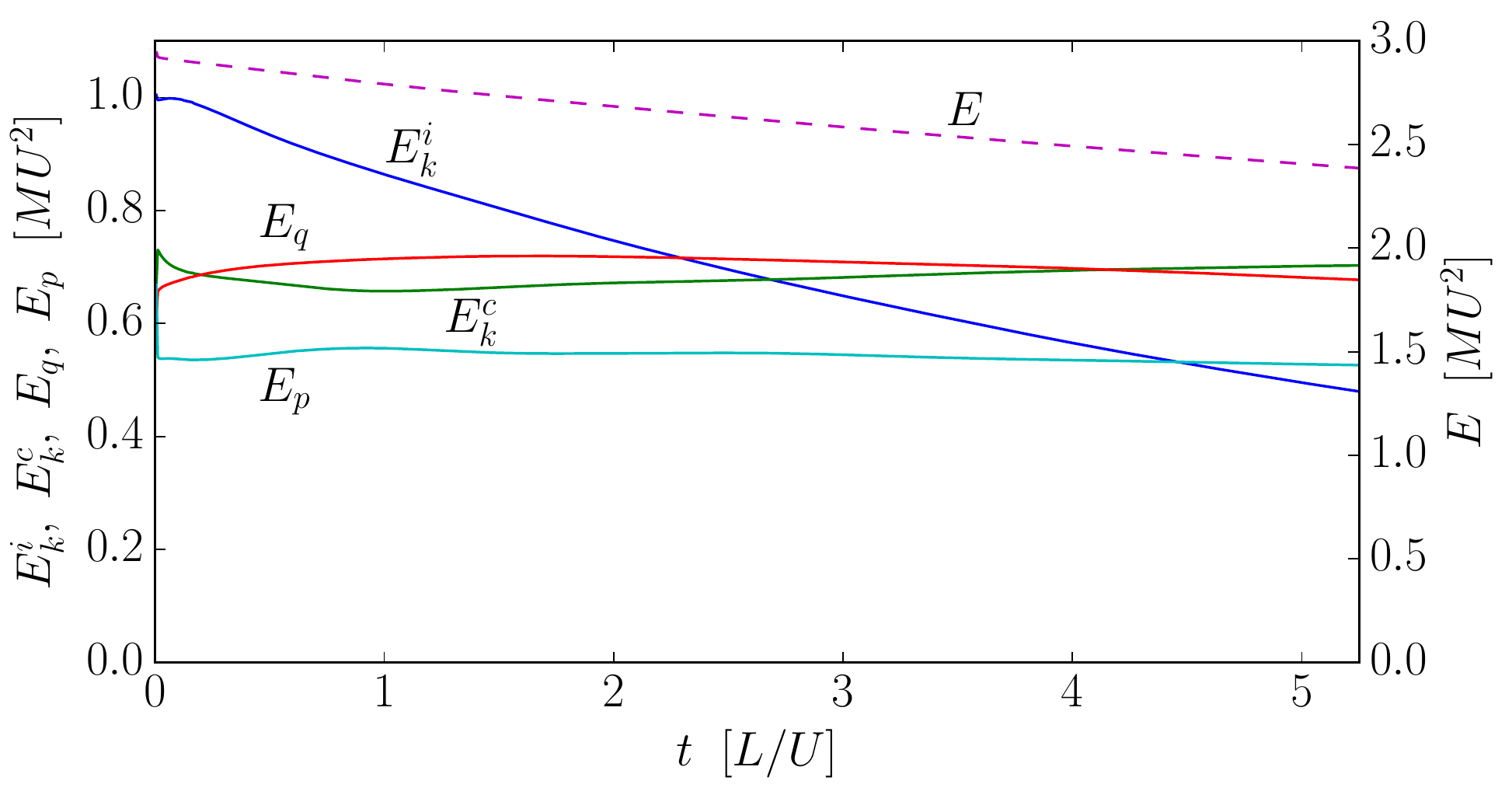}
    \caption{{\it (Color online)} Evolution of the total energy and its
    different components for the simulation with $T=0.64 T_\lambda$ and
    $N=4096$.}
    \label{evoltot}
\end{figure}

In order to characterize the system we first perform a temperature
scan solving the SGLE with the chemical potential $\mu$ adjusted to keep
the total density $\rho_0=1$. In Fig.~\ref{condf} we show the condensate
fraction $\mathcal{N}_0/\langle \mathcal{N} \rangle$  (with
$\mathcal{N}_0 = |A_0|^2$) at late times in the evolution, as a
function of the temperature $T$. As reported before in
\citet{Krstulovic11a,Krstulovic11b}, the typical behavior for second
order transitions can be observed, with 
$\mathcal{N}_0/\langle \mathcal{N} \rangle \approx 0$ for $T>T_\lambda$, and 
$\mathcal{N}_0/\langle \mathcal{N} \rangle$ growing as in a phase transition for
$T<T_\lambda$. The value of the critical temperature $T_\lambda$ was
determined from this analysis. The scans were performed at two
different linear resolutions $N=128$ and $N=1024$. The results from
both coincide, showing the simulations are well converged. Also shown in
the figure is the usual prediction for the condensate fraction coming
from ideal BEC theory \cite{Pathria} where $\mathcal{N}_0/\langle \mathcal{N}
\rangle = 1 - (T/T_\lambda)^{3/2}$. This prediction does not match our
results exactly as it is derived for non-interacting bosons, which is
not our case. Nonetheless, the behaviors are similar. 

As explained above, these thermal states were coupled to solutions of
the ARGL to generate initial conditions for the GPE at different
temperatures. In Fig.~\ref{massk} we show the spectrum of the mass
fluctuations $\rho(k)$ of the initial condition for five different
temperatures, as well as the incompressible kinetic energy spectrum
$E_k^i(k)$, and the compressible kinetic energy spectrum $E_k^c(k)$. In
all cases, the increasing amplitude of high wavenumber (small scale)
modes as $T$ is increased (but specially in $E_k^c(k)$, associated with
phonon excitations) accounts for the  increasing thermal effects. Note
however that the low wavenumber (large scale) spectrum of $E_k^i(k)$,
associated with the initial ABC flow, remains largely unaffected by the
thermal fluctuations, a result of the sufficient scale separation in
these runs.

\begin{figure*}
    \centering
    \includegraphics[width=0.9\textwidth]{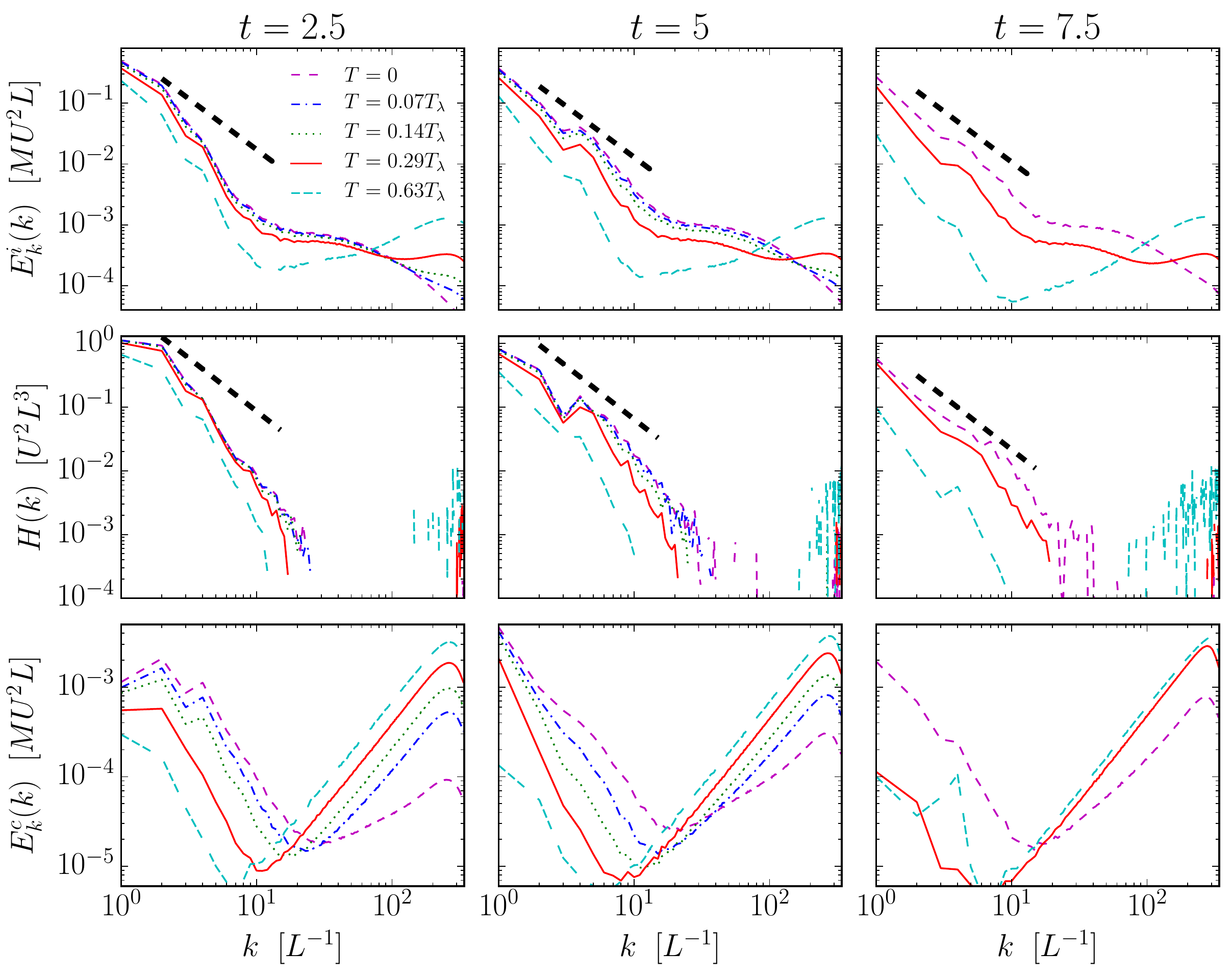}
    \caption{{\it (Color online)} Spectra of incompressible kinetic
      energy, helicity, and compressible kinetic energy at different
      temperatures and at different times. All simulations have
      $N=1024$ linear resolution. The thick dashed line indicates
      $k^{-5/3}$ scaling, predicted for the dual cascade of energy and
      helicity \cite{Brissaud73}, and previously observed in zero
      temperature simulations \cite{Clark17}. This scaling is
      compatible with the simulations at at low temperatures, but it
      is lost once the temperature is increased and viscous effects
      become strong enough.}
    \label{specscan}
\end{figure*}

\begin{figure*}
    \centering
    \includegraphics[width=0.9\textwidth]{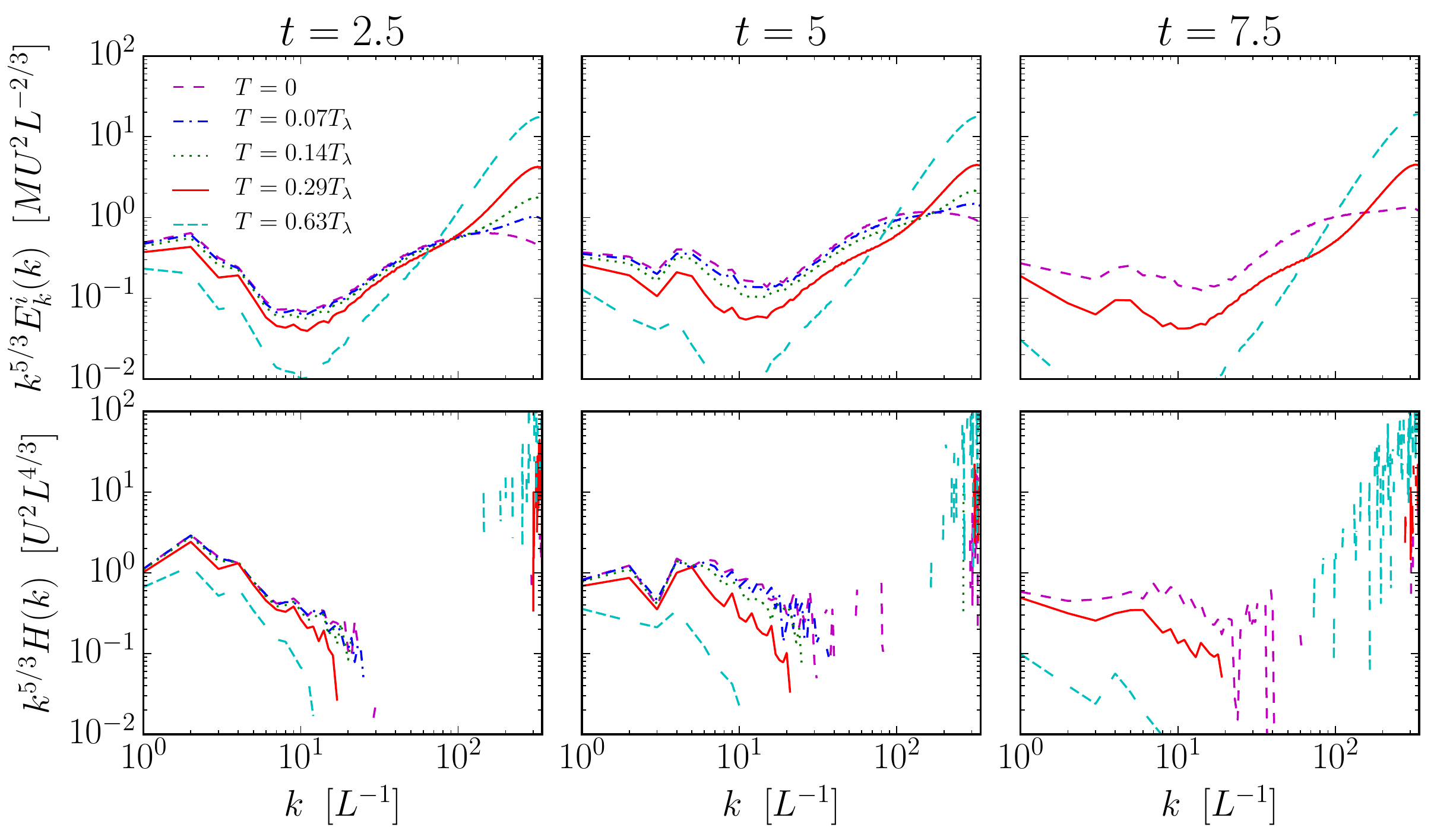}
    \caption{{\it (Color online)} Compensated version of the
    incompressible kinetic energy and helicity spectra shown in
    Fig.~\ref{specscan}. As seen in the figure, only at the smaller
    temperatures some range of wave numbers compatible with 
    $k^{-5/3}$ scaling (i.e., an approximately flat compensated
    spectrum) is recovered.}
    \label{compensated}
\end{figure*}

\subsection{Dynamical evolution} 

We now focus on understanding finite temperature effects on the
evolution of the GPE. We thus show results from six different
simulations. Five of them were done at a linear resolution of
$N=1024$, with temperatures ranging from zero to $T=0.63T_\lambda$, 
while the sixth simulation was performed at a linear resolution of
$N=4096$ and at $T=0.64T_\lambda$.

\subsubsection{Large-scale flow structure} 

We begin by showing two visualizations of the density field for the
simulation with $N=4096$ and $T=0.64T_\lambda$ at time $t\approx1$. In
Fig.~\ref{fullbox_dvr}, a volumetric rendering of mass density is
shown using VAPOR \cite{Clyne07}. Similarly to the zero temperature
quantum ABC flow \cite{Clark17}, large vortex bundles are formed
within the flow, and regions of quiescence (with almost no vorticity)
appear. At the larger scales the structure of the flow looks similar
to that of a classical ABC flow, as expected. Moreover, although the
thermal fluctuations blur the small scales, the large scale flow is
clearly discernible. In Fig.~\ref{fullbox_iso} isosurfaces of the
density field are shown.  As in Fig.~\ref{fullbox_dvr}, and contrary
to the zero temperature case where it is easy to spot individual
vortices (see \cite{Clark17}), the thermal noise lumps the vortices
inside the bundles, making it difficult to discern individual
structures from visual inspection, although traces of their presence
are evident.

To further confirm the coexistence of large-scale correlations
(associated with the flow) with small-scale thermal fluctuations and
vortices, we show in Fig.~\ref{fullbox_corr} the spatial correlation
function of mass density fluctuations 
\begin{equation}
  C(d) = \left< (\rho({\bf x} +d\hat{x}) - \rho_0)(\rho({\bf x}) -
    \rho_0) \right> / \rho_0^2 ,
\end{equation}
for the simulation with $N=4096$ and $T=0.64T_\lambda$ at time
$t\approx1$, where $d$ is the spatial displacement (which in
Fig.~\ref{fullbox_corr} is normalized in units of the healing length
$\xi$). The function $C(d)$ is also proportional, by the
Wiener-Khinchin theorem, to the Fourier transform of the internal
energy spectrum. Note $C(d)$ decays rapidly with $d/\xi$ in a distance 
proportional to the vortex core size, thus further confirming the
presence of quantized vortices in the flow. Then, $C(d)$ remains almost
constant up to very long-range distances ($d\approx 10^3 \xi$),
confirming the presence of a large-scale structure in the system.

\subsubsection{Energy and helicity decay} 

In the zero temperature case nonlinear interactions of Kelvin waves
lead to the emission of phonons \cite{Lvov10,Vinen02}, which deplete
the incompressible kinetic energy \cite{Nore97a} and the helicity
\cite{Clark17}. The presence of thermal noise adds a new depletion
mechanism. In order to study it, we show in Figs.~\ref{eht} and
\ref{ehtlog} the evolution of the incompressible kinetic energy
$E^i_k$ and of the helicity $H$ for five different temperatures, in
linear and in semi-logarithmic scales respectively.

As expected, for all temperatures both $E^i_k$ and $H$ decay in time. At
very early times a short transient can be seen (due to the system
correcting frustration effects coming from the initial conditions),
after which the different dynamical mechanisms come into play. This
transient is similar for all the runs, and almost independent of the
temperature. After this transient, at low temperatures both the
incompressible energy and the helicity decay very slowly or remain
approximately constant (see in particular the case with $T=0$ in
Fig.~\ref{ehtlog}), up until $t \approx 3$. This is similar to what is
observed in freely decaying classical turbulence: in that case the early
``inviscid-like'' phase corresponds to the build up of the turbulent
cascade while dissipation remains negligible, and which (in the
classical case) ends when small scale excitations reach the viscous
dissipation scale. In classical turbulent flows, the presence of
helicity is known to extend the duration of this ``inviscid-like'' phase
(see, e.g., \cite{Teitelbaum09} and references therein). As explained in
\citep{Clark17}, in the quantum case and for $T$ close to zero this
inviscid phase corresponds to the time during which vortices interact
and the Kelvin wave cascade builds up, so after $t\approx 4$ the
emission of phonons becomes prominent and the incompressible kinetic
energy and the helicity start being depleted. Note that during this
phase both energy and helicity are transferred towards smallers scales,
as will be confirmed later by the energy and helicity spectra. 

Unlike the simulations at low temperature, the simulations at the
highest temperatures go directly from the short initial transient to a
seemingly exponential decay, without an inviscid-like phase in between
(see Fig.~\ref{ehtlog}). At late times ($t>6$), all simulations show
similar exponential decay rates (see Fig.~\ref{ehtlog}) as a significant
fraction of the energy has already thermalized, with the exception of
the simulation with $T=0.64 T_\lambda$ and $N=1024$, which has a higher
initial temperature and thus can reach a thermal equilibrum faster.  The
exponential decay observed in these runs is reminiscent of what is
observed in the free decay of low Reynolds classical flows. To further
verify this we show estimations of the exponential decay rates
$-(1/E^i_k) d E^i_k /dt$ and $-(1/H)dH/dt$ in Fig.~\ref{exponential}.
For the higher temperatures these magnitudes become close to constant
for long periods of time, confirming an exponential decay. This is not
the case for the lower temperatures, where oscillations and a late
growth of these quantities are present at all times. Moreover, the
exponential decay behavior at high temperatures is compatible with weak
nonlinearities and can be used, as explained below, to estimate an
effective eddy viscosity of the flow by assuming a governing equation
for the velocity of the Stokes form, $\partial{\bf v}/\partial t \approx
\nu_\textrm{eff}\nabla^2{\bf v}$.

\begin{figure}
    \centering
    \includegraphics[width=8.5cm]{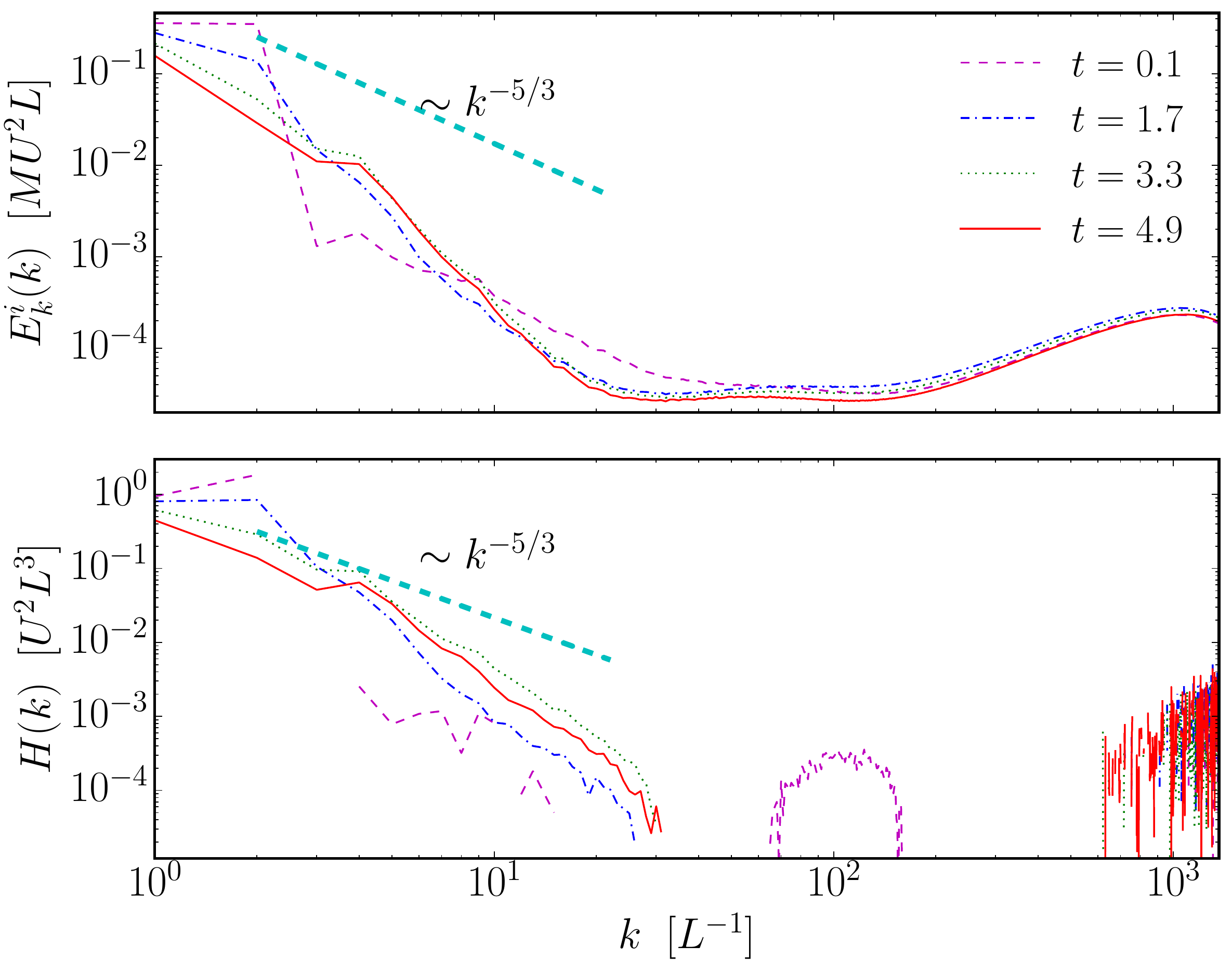}
    \caption{{\it (Color online)} Spectra of the incompressible
      kinetic energy and of the helicity at $T=0.64T_\lambda$ in the
      simulation with $N=4096$ at different times. A turbulent scaling
      law is shown as a reference. Although nonlinear excitations
      develop, the range of scales compatible with the turbulent
      scaling is short and at late times the spectrum decays rapidly.}
    \label{spectra}
\end{figure}

So far, these results indicate several things: The effects of the
thermal states generated with the SGLE upon the quantum turbulent flow
can be modeled, at least for global quantities and in the simplest
scenario, using an effective viscous dissipation. This effect can be,
at the highest temperatures considered, strong enough (even at the
highest resolution) that nonlinear interactions and Kelvin wave
turbulence cannot fully develop, such that (pseudo) viscous effects
dominate the dynamics. As observed from Fig.~\ref{exponential}, the
rate of change of the energy at zero temperature is almost negligible
at $t\approx 1$, but increases and becomes considerable in the other
cases. Thus, we can draw from this fact to construct anzats for the
effective viscosity as a function of temperature. In a freely decaying
classical flow an eddy viscosity can be estimated as 
$\nu_\textrm{eff} = -(\tilde{L}^2/E^i_k) dE^i_k/dt$, where $\tilde{L}$ is some
large-scale correlation length. Here we have several choices for a
characteristic length $\tilde{L}$: a fixed length $L_0$ given by the length
scale of the large-scale flow at $t=0$, the integral scale (i.e., the
correlation length) of the incompressible velocity field
\begin{equation}
L_i = \frac{2\pi}{E_k^i} \int{ \frac{E_k^i(k)}{k} dk } ,
\end{equation}
(where $E_k^i(k)$ is the spectrum of the incompressible kinetic
energy), the intervortex distance $\ell$, or the healing length
$\xi$. We verified that the behavior with temperature of
$\nu_\textrm{eff}$ with all these choices for $\tilde{L}$ is qualitatively
similar, except for a prefactor, and thus show in Fig.~\ref{scaling}
two estimations of $\nu_\textrm{eff}$ based on large-scale correlation
lengths: the fixed length $L_0$ and the integral length $L_i$. The
viscosity estimates are close to zero for $T=0$, grow linearly with
temperature up to $T/T_\lambda \approx 0.3$, and then either keep
growing at a lower rate or decrease for larger temperatures, depending
on the choice of $L$. Moreover, the estimations of  $\nu_\textrm{eff}$
for the $N=1024$ and the $4096$ runs at the highest temperature are
similar.

These results can be interpreted as follows. The viscosity of the
normal fluid $\nu_{n}$ can be expected to be proportional to mean free
path $\lambda_m$ times the sound velocity, i.e., 
$\nu_{n} \sim \lambda_m c$. When we increase the resolution fixing 
$\xi k_{\rm max}$ (as done here), $c$, and the temperature $T$, the
mean free path depending only on the temperature is constant (in units
of $\xi$), i.e., 
\begin{equation}
\lambda_m \sim \xi f(T/T_\lambda) ,
\end{equation}
where $f(T/T_\lambda)$ is a dimensionless function. Therefore
$\nu_{n}$ should scale as the inverse of the spatial resolution,
$\nu_{n} \sim 1/N$. But this argument holds as long as the mean free
path is smaller than the box size, $\lambda_m<2 \pi$, while the mean
free path diverges when $T\to 0$. Thus, at a given temperature the
viscosity of the normal fluid should first remain constant with
resolution, and then after a certain critical resolution go to zero as
$1/N$. This is for the normal fluid alone, and its contribution to the
total flow should scale as $\rho_n/\rho \sim T$. Thus, we can expect
an effective viscosity measured on the total fluid to scale as
\begin{equation}
\nu_{\rm eff}\sim \nu_{n} \rho_n/\rho \sim \nu_{n} T/T_\lambda ,
\end{equation}
which should first grow like $T$ and then decrease when the mean free
path becomes less than the box size. Further confirmation of this
scaling would require a direct measure of the mean free path; we
discuss possible methods to achieve this goal in the conclusions.

Finally, it is important to note that, as in the zero temperature case,
the Galerking truncated GPE conserves the total energy, and that our
spatial discretization method is also conservative (although time
discretization introduces new errors as discussed next). So, while the
incompressible kinetic energy is depleted, the other components of the
energy can be expected to grow. As an illustration, the evolution of the
total energy, the incompressible kinetic energy, the compressible
kinetic energy, the quantum energy, and the potential energy for the
simulation with $N=4096$ and $T=T0.64T_\lambda$ is shown in
Fig.~\ref{evoltot}. Note a fraction of the total energy is indeed lost
due to numerical errors, resulting from the fact that the great cost of
doing such a high resolution simulation did not allow us to use a very
small time step.  Nonetheless, energy is conserved up to 95\% when
$t\approx1$ (which is when most of the physics we are interested occurs)
and up to 82\% at the very end of the simulation.

\subsubsection{Spatial spectra}

Finally, we study the effect of temperature on the evolution of the
spatial spectra of the two components of the kinetic energy
(compressible and incompressible), and of the helicity. This should give
further confirmation that for large enough temperatures, the nonlinear
cascade of energy and of helicity is strongly arrested. The results for
the simulations with $N=1024$ are shown in Fig.~\ref{specscan} (with 
compensated versions of the spectra shown in Fig.~\ref{compensated}),
while the results for the simulation with $N=4096$ are shown in
Fig.~\ref{spectra}. Note that the compensated spectra stemming from
the simulations with $N=1024$ shown in Fig.~\ref{compensated} are 
expected to be flat in regions that follow Kolmogorov-like scaling;
animations showing the evolution of each spectra can be also found in 
\cite{SI}.

While, as shown in Fig.~\ref{massk}, the initial spectra at small
wavenumbers (large scales) are relatively similar for all temperatures,
differences can already be seen in Fig.~\ref{specscan} at $t=2.5$ for
the simulations at different temperatures. In particular, the
simulations with the largest temperatures have less power at small
wavenumbers (for all quantities $E_k^i$, $H$, and $E_k^c$), and more
power at large wavenumbers (specially for $E_k^c$), as can be expected
from the larger thermal fluctuations. As the flow evolves and nonlinear
interactions take place (see the spectra at $t=5$), the low temperature
simulations develop a range of wavenumbers compatible with a
Frisch-Brissaud dual cascade of energy and of helicity towards small
scales \cite{Brissaud73} (which corresponds to Kolmogorov-like scaling
for both spectra), previously observed in zero temperature simulations
\cite{Clark17}. However, the simulations with the highest temperatures
do not develop a broad spectrum, and although excitations grow at
intermediate and at small wavenumbers in $E_k^i(k)$ and $H(k)$, 
the spectra drops faster confirming the effect of damping discussed in
the previous section, and in agreement with the effect expected for a
large effective viscosity. The compensated spectra shown in
Fig.~\ref{compensated} confirm this. The $N=4096$ simulation (see
Fig.~\ref{spectra}) also shows this damped behavior, and as estimated
from the results in Fig.~\ref{scaling} has an effective viscosity of the
same order as the simulation with $N=1024$ at a similar temperature.

All spectra at all temperatures have a pronounced change (or knee) at
around $k\approx10$. At low temperatures, this bump (which in the case
of the spectrum of $E_k^i$ is followed by a range of wavenumbers with
decreasing amplitude as $k$ increases) can be associated with a
bottleneck produced by the Kelvin wave cascade at scales smaller than
the mean intervortex distance \cite{Clark17}. This is seen more clearly
in the compressible kinetic energy spectra. For the simulations at the
highest temperatures this second range is swallowed up by the presence
of the thermalized modes. As it can be expected, the flat portion of the
spectra between the cascading part at small wavenumers and the
thermalized part at large wavenumbers is wider in the $N=4096$
(Fig.~\ref{spectra}) simulation compared to the ones at $N=1024$
(Fig.~\ref{specscan}). The spectra of the helicity fluctuate around zero
with fast changes in sign above this wavenumber in all cases, a result
of the depletion of helicity by phonons (as the spectra are plotted in
logarithmic scale only positive values are shown, the missing parts
correspond to the negative values). The spectrum of compressible kinetic
energy grows as $\sim k^2$, which can be expected for a thermalized
state (and its amplitude increases with increasing temperature).

\section{Conclusions} 
\label{conclusions}

Modeling quantum flows at nonzero temperatures is key to understand
recent experimental results of quantum turbulence. However, models for
quantum flows at finite temperature are limited, sometimes derived from
phenomenological models, in other cases obtained from coarse
approximations, and in many cases their dynamics have not been fully
characterized. Here we presented a study of helical quantum turbulence
at various temperatures using very large resolution simulations and a
model based on the Gross-Pitaevskii equation with thermal states
generated by the Stochastic Guinzburg-Landau equation.

Our results show that in this model, under the presence of thermal
noise, a quantum flow can behave as a viscous classical flow, with
exponential decay of the incompressible kinetic energy and of the
helicity. A smooth transition between the behavior at zero
temperatures and at large temperatures (for temperatures lower than
the critical) was reported. Moreover, the (pseudo) viscous effects can
strongly quench the formation of a turbulent cascade, even at the
largest spatial resolution considered. However, when the temperature
is not too high, a dual cascade of energy and of helicity (as also
observed in classical turbulence and in quantum flows at zero
temperature) can be reobtained.

We presented a phenomenological estimation of the effective viscosity
in this model, which shows linear scaling with increasing temperature,
and a saturation for very high temperatures. An argument based on the
mean free path accounts for this behavior, and opens the door to
better estimations of the effective viscosity by measuring directly
this lengthscale. This can be done by studying the spatio-temporal
spectrum of the flow as a function of the temperature, which gives
access to the spectrum of phonons in the system
\cite{Clark15a}. However, as computation of this spectrum is
computationally intensive, it can only be done at lower resolutions or
using a different flow configuration, and is thus left for future
work.

\begin{acknowledgements}
The authors acknowledge financial support from Grant No.~ECOS-Sud
A13E01, and computing hours in the CURIE supercomputer granted by
Project TGCC-GENCI No.~T20162A711. P.C.dL. acknowledges funding from the
European Research Council under the European Community's Seventh
Framework Program, ERC Grant Agreement No.~339032.
\end{acknowledgements}

\bibliography{ms}

\end{document}